\documentclass[aps,pra,onecolumn,reprint,superscriptaddress,longbibliography]{revtex4-2}

\usepackage{hyperref}
\usepackage{graphicx}  
\usepackage{bm}        
\usepackage{amssymb}   
\usepackage{xspace}
\usepackage{verbatim}
\usepackage{color}
\usepackage{soul}
\usepackage{subfigure}
\usepackage[export]{adjustbox}
\usepackage{array}
\usepackage{dcolumn}
\usepackage{amsthm,stmaryrd,mathtools}
\usepackage{txfonts}
\usepackage{braket}
\usepackage{amsmath}
\usepackage{xcolor}
\usepackage{array}
\usepackage{multirow}
\usepackage{tabularx}
\usepackage{booktabs} 
\usepackage{lipsum}
\usepackage{dsfont}
\definecolor{lightblue}{RGB}{100,100,250}

\begin{document}

\global\long\def\ket#1{\left|#1\right\rangle }%
\global\long\def\bra#1{\left\langle #1\right|}%
\global\long\def\braket#1#2{\langle#1|#2\rangle}%
\global\long\def\expectation#1#2#3{\langle#1|#2|#3\rangle}%
\global\long\def\average#1{\langle#1\rangle}%

\author{Rozhin Yousefjani}%
\email{ryousefjani@hbku.edu.qa}
\affiliation{Qatar Center for Quantum Computing, College of Science and Engineering, Hamad Bin Khalifa University, Doha, Qatar}
\author{Saif Al-Kuwari}%
\affiliation{Qatar Center for Quantum Computing, College of Science and Engineering, Hamad Bin Khalifa University, Doha, Qatar}

\title{Exponentially-enhanced Weak-field Sensing with Quantum Stark Localization}

\begin{abstract}
Stark-localized quantum probes have recently been shown to enable quantum-enhanced weak-field sensing with polynomial or super-polynomial scaling. In this paper, we show that the spatial geography of the encoded field can elevate this advantage to a genuine exponential scaling. We study a one-dimensional Stark probe subject to an exponential gradient profile, \(V_j=e^{aj}\), and analyze its metrological performance in both equilibrium and non-equilibrium regimes, for single-particle and interacting many-body settings. In the equilibrium single-particle case, we derive an analytical lower bound showing that the quantum Fisher information grows exponentially with system size, and confirm numerically that this enhancement persists throughout the extended phase and at the localization transition. We further show that the same exponential scaling survives for mid-spectrum eigenstates and in the interacting many-body regime. This advantage remains intact under a fair resource analysis because the relevant preparation gap closes only algebraically, so the polynomial preparation overhead cannot offset the exponential gain in sensitivity. In the non-equilibrium regime, a simple product-state initialization followed by free evolution already retains exponential enhancement, eliminating the need for cooling, adiabatic preparation, or operation within a narrowly tuned sensing window. Finally, we outline a superconducting implementation based on flux-tunable transmon qubits with graded mutual inductive coupling to a common sensing bus. Our results identify exponentially graded Stark potentials as a distinct and experimentally plausible route to weak-field sensing with exponentially improving precision.
\end{abstract}

\maketitle

\section{Introduction}

Quantum sensing exploits uniquely quantum features to estimate weak fields and couplings with a precision beyond classical limits. While early advances were primarily based on interferometric protocols and specially prepared entangled states \cite{Giovannetti2004,Paris2009,giovannetti2011advances,degen2017quantum,yousefjani2017enhancement,beau2017nonlinear}, it is now well established that quantum many-body systems provide a broader route to enhanced metrology \cite{Montenegro2025Review}. In such probes, the sensing advantage is not tied to a single mechanism, but can emerge from collective many-body phenomena that strongly amplify the response to weak perturbations, including 
first-order phase transitions~\cite{Raghunandan2018HighDensity,Mirkhalaf2020Supersensitive,Heugel2019QuantumTransducer,sarkar2025exponentially}, 
second-order criticality~\cite{Zanardi2006GroundStateOverlap,Zanardi2008QuantumCriticality,Rams2018LimitsCriticalityMetrology,Chu2021DynamicFramework,Liu2021ExperimentalCritical,Montenegro2021GlobalSensing,Wald2020InOutEquilibrium,Mishra2021DrivingEnhanced}, 
Floquet-engineered critical phenomena~\cite{Montenegro2023BoundaryTimeCrystals}, 
time-crystalline phase~\cite{Iemini2024FloquetTimeCrystals,Yousefjani2025DiscreteTimeCrystal,Montenegro2023BoundaryTimeCrystals}, 
Stark and quasiperiodic localization~\cite{He2023PRL,Sahoo2024LocalizationDriven,Yousefjani2023CPB,Yousefjani2025PRApplied}, 
and topological phase transitions~\cite{Sarkar2022FreeFermionic,Sarkar2024CriticalNonHermitian}.
This broader perspective has substantially expanded the scope of quantum metrology, showing that both equilibrium and dynamical many-body physics can be systematically harnessed for parameter estimation \cite{Montenegro2025Review,Abiuso2025FundamentalLimits}.
\\ \\
Within this broader program, Stark localization has emerged as a promising setting for weak-field sensing. Unlike disorder-induced localization~\cite{monteiro2021quantum,yousefjani2023mobility,yousefjani2023floquet,sajid2025thermal}, it is generated by a deterministic field gradient, which makes the underlying mechanism cleaner and the sensing geometry directly controllable. Previous work established that Stark-localized probes can achieve quantum-enhanced sensitivity in both single-particle and interacting many-body settings. In the extended regime, the precision grows superlinearly with system size, whereas deep in the localized regime this enhancement disappears and the sensitivity approaches a nearly size-independent behavior \cite{He2023PRL,Yousefjani2023CPB}. More recently, it was shown that nonlinear gradient profiles can substantially reshape this metrological response, making clear that the performance of a Stark probe is governed not only by the field amplitude, but also by the spatial profile through which the field is encoded \cite{Yousefjani2025PRApplied}. This raises a sharper question that has not been addressed so far. Can field geography alone drive Stark sensing precision beyond the polynomial and super-polynomial regimes identified for linear and power-law profiles?
\\ \\ 
In this work, we answer this question by considering an exponential Stark profile, $V_j=e^{aj}$, and analyzing its sensing performance in both equilibrium and non-equilibrium regimes, for single-particle and interacting many-body probes. We show that this change in field geography qualitatively alters the metrological scaling, promoting the sensitivity from the polynomial or super-polynomial behavior known for linear and power-law Stark probes to a genuine exponential growth with system size. In the equilibrium single-particle setting, we derive an analytical lower bound demonstrating exponential growth of the quantum Fisher information and confirm numerically that this enhancement persists throughout the extended phase and at the localization transition. We further show that the same qualitative behavior survives away from the ground state, with mid-spectrum eigenstates displaying exponential sensitivity as well, and that it remains robust in the interacting many-body regime. Importantly, this advantage survives a fair resource analysis. Although state preparation in the equilibrium protocol requires adiabatic evolution, the relevant gap closes only algebraically, so the polynomial preparation overhead cannot overcome the exponential gain in precision. Finally, in the non-equilibrium regime, we show that a simple product-state initialization followed by free evolution already retains exponential enhancement, thereby eliminating the need for cooling, adiabatic preparation, or operation in a narrowly tuned sensing window.
\\ \\
These results also clarify that exponentially enhanced quantum sensing can arise through physically distinct mechanisms. In recently studied first-order critical sensors, the exponential advantage is tied to an exponentially closing energy gap at criticality~\cite{sarkar2025exponentially}. In the present Stark setting, by contrast, the enhancement has a conceptually different root and is driven by the exponential spatial geography of the encoded field itself rather than by an exponentially small preparation gap. In fact, the field profile acts as the key metrological resource. Beyond the scaling analysis, we also outline a superconducting implementation based on flux-tunable transmon qubits coupled to a common sensing bus with graded mutual inductances, which offers an experimentally possible route to realizing the exponential Stark coupling while maintaining a clean separation between the engineered spatial response and the unknown signal amplitude.

\section{Quantum Parameter Estimation}
In estimating an unknown parameter $h$ encoded in a quantum probe 
state $\rho(h)$, any measurement described by a POVM $\{\Pi_k\}$ 
yields outcomes distributed according to $p_k(h) = 
\mathrm{Tr}[\Pi_k \rho(h)]$. The precision of such a measurement, 
quantified by the variance $\delta h^2 = \langle h^2 \rangle - 
\langle h \rangle^2$, is bounded from below by the 
Cram\'{e}r--Rao inequality
\begin{equation}
    \delta h^2 \geq \frac{1}{M \mathcal{F}_C(h)},
\end{equation}
where $M$ is the number of independent repetitions and 
$\mathcal{F}_C(h) = \sum_k p_k(h)\left(\partial_h \ln 
p_k(h)\right)^2$ is the Classical Fisher Information 
(CFI)~\cite{Fisher1922,cramer1999mathematical,Paris2009}. Maximizing the CFI over 
all possible measurements yields the Quantum Fisher Information 
(QFI), $\mathcal{F}_Q(h) = \max_{\{\Pi_k\}} \mathcal{F}_C(h)$, 
and the quantum Cram\'{e}r--Rao bound
\begin{equation}
    \delta h^2 \geq \frac{1}{M \mathcal{F}_Q(h)}.
    \label{eq:QCRB}
\end{equation}
For a pure probe state $\rho(h) = |\psi(h)\rangle\langle\psi(h)|$, 
the QFI reduces to the closed form~\cite{braunstein1994statistical,Paris2009,rao1992information}
\begin{equation}
    \mathcal{F}_Q(h) = 4\left[\langle \partial_h \psi | 
    \partial_h \psi \rangle - \left|\langle \partial_h \psi | 
    \psi \rangle\right|^2\right].
    \label{eq:QFI_pure}
\end{equation}
The scaling of $\mathcal{F}_Q$ with the probe size $L$ sets the 
metrological performance class, $\mathcal{F}_Q \propto L$ defines 
the standard quantum limit, whereas any super-linear or exponential 
growth, $\mathcal{F}_Q \propto L^\alpha$ with $\alpha > 1$, or 
$\mathcal{F}_Q \propto e^{\beta L}$, constitutes genuine quantum 
enhancement~\cite{Giovannetti2004,Montenegro2025Review}.

For quantum sensors based on strongly correlated many-body systems, 
the Hamiltonian takes the form $H = H_C + h H_0$, where $H_C$ and 
$H_0$ are competing terms with $[H_C, H_0] \neq 0$. When these 
terms are comparable in magnitude, the system may undergo a phase 
transition, which has proven to be a powerful resource for achieving 
quantum-enhanced sensitivity~\cite{Montenegro2025Review}. For equilibrium probes initialized in 
an eigenstate $|E_l\rangle$ of $H$, the QFI is equivalent to the 
fidelity susceptibility~\cite{You2007} and takes the form
\begin{equation}
    \mathcal{F}_Q = 4 \sum_{k \neq l} 
    \frac{|\langle E_k | H_0 | E_l \rangle|^2}
         {(E_k - E_l)^2},
    \label{eq:QFI_eigenbasis}
\end{equation}
where $\{E_k\}$ and $\{|E_k\rangle\}$ are the eigenvalues and 
eigenstates of $H$. Achieving quantum-enhanced sensitivity in this 
setting requires preparing the probe in a specific eigenstate, 
typically the ground state, in the vicinity of the phase transition 
point. This places two practical constraints on the protocol. First, 
the adiabatic preparation time is inversely proportional to the 
minimum energy gap $\Delta$ of the Hamiltonian, $\tau \sim 1/\Delta$, 
which must be accounted for in any fair resource analysis. Second, the enhancement is generically confined to a narrow region around 
the critical point, requiring substantial prior knowledge of $h$ 
and making the protocol unsuitable for global sensing tasks.

These constraints motivate the use of non-equilibrium dynamics as 
an alternative sensing resource. In this approach, the probe is 
initialized in a simple product state, and the unknown parameter is 
encoded through the unitary time evolution $|\psi(t, h)\rangle = 
e^{-iHt}|\psi(0)\rangle$. For a closed system evolving from a pure 
initial state, the time-dependent QFI is given by
\begin{equation}
    \mathcal{F}_Q(t, h) = 4\left[\langle \partial_h \psi(t,h) | 
    \partial_h \psi(t,h) \rangle - \left|\langle \partial_h 
    \psi(t,h) | \psi(t,h) \rangle\right|^2\right].
    \label{eq:QFI_dynamic}
\end{equation}
In non-equilibrium probes, time enters as an independent resource 
alongside the probe size $L$. The standard quantum limit in this 
setting corresponds to $\mathcal{F}_Q \propto tL$, and quantum 
enhancement is identified by any scaling that exceeds this 
baseline in either resource.

\section{Gradient field sensing; Equilibrium probe}
Stark-based probes have recently emerged as a prominent platform for measuring power-law magnetic-field gradients, with demonstrations spanning both single-particle settings and interacting many-body systems~\cite{He2023PRL,Yousefjani2025PRApplied,Yousefjani2023CPB}.
Here, we take this further and ask how this landscape changes when the gradient potential is no longer power-law. Specifically, we explore how the shape of the potential modifies the sensing capabilities of Stark probes in both single-particle and many-body interacting regimes.

\subsection{Single-particle probe}
\begin{figure}[t!]
    \centering
    \includegraphics[width=\linewidth]{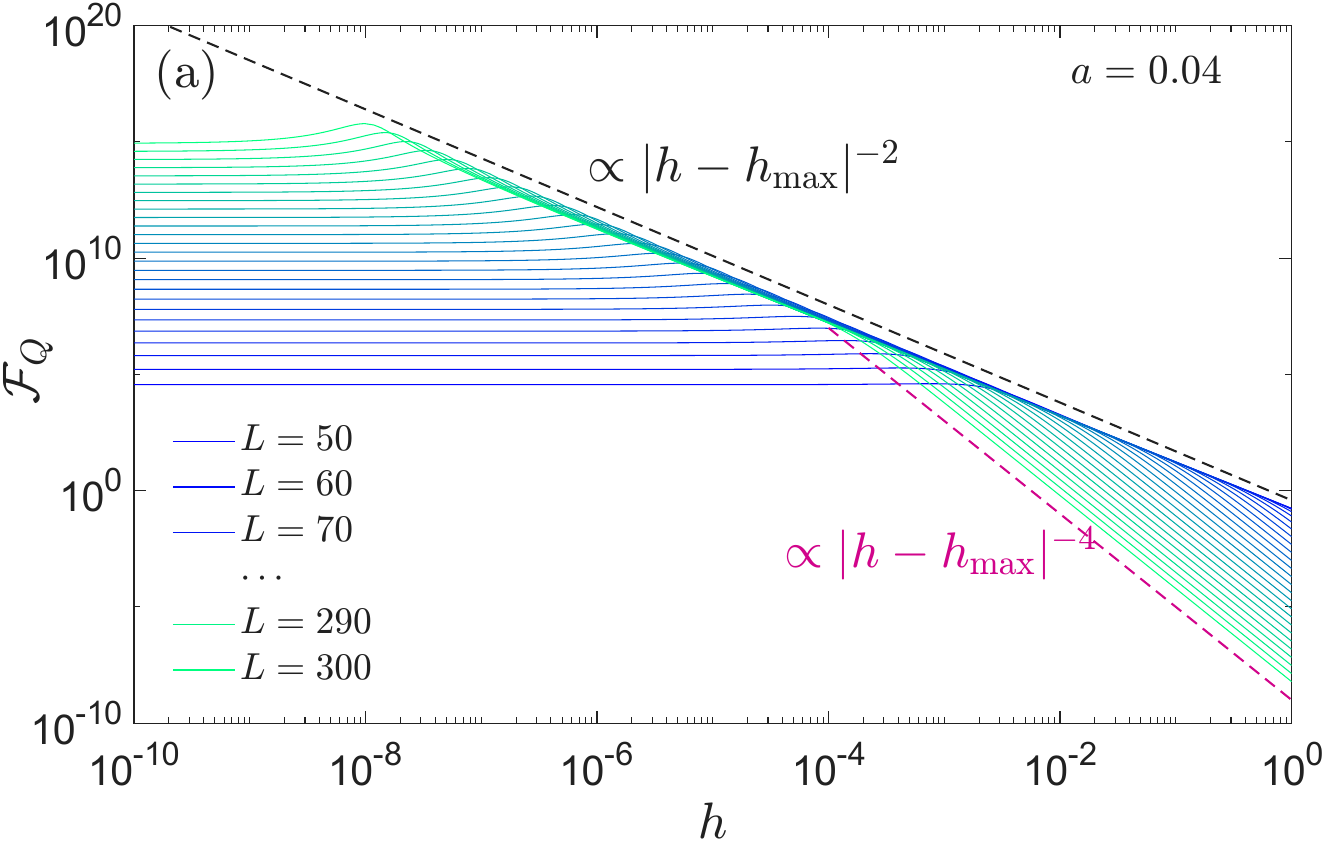}
    \includegraphics[width=\linewidth]{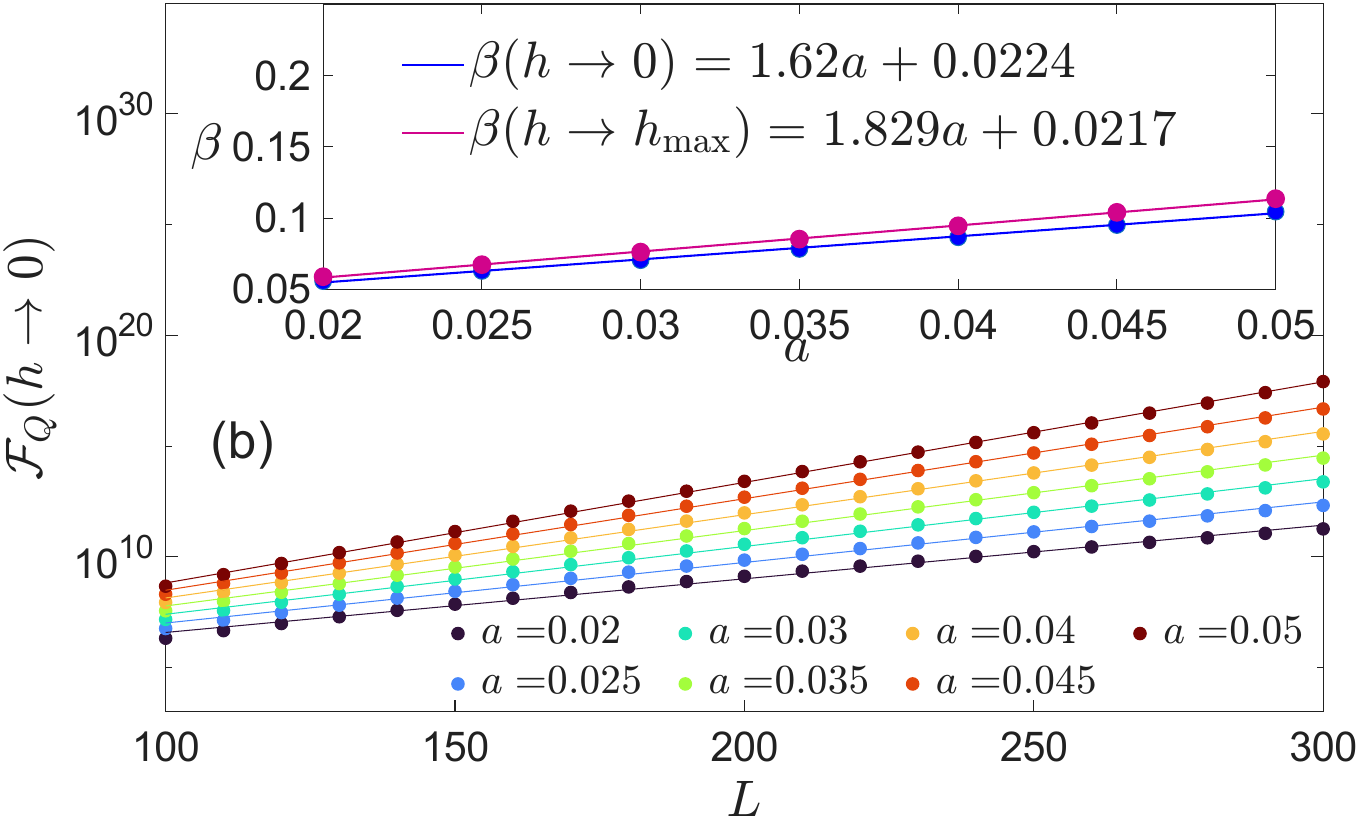}
    \includegraphics[width=\linewidth]{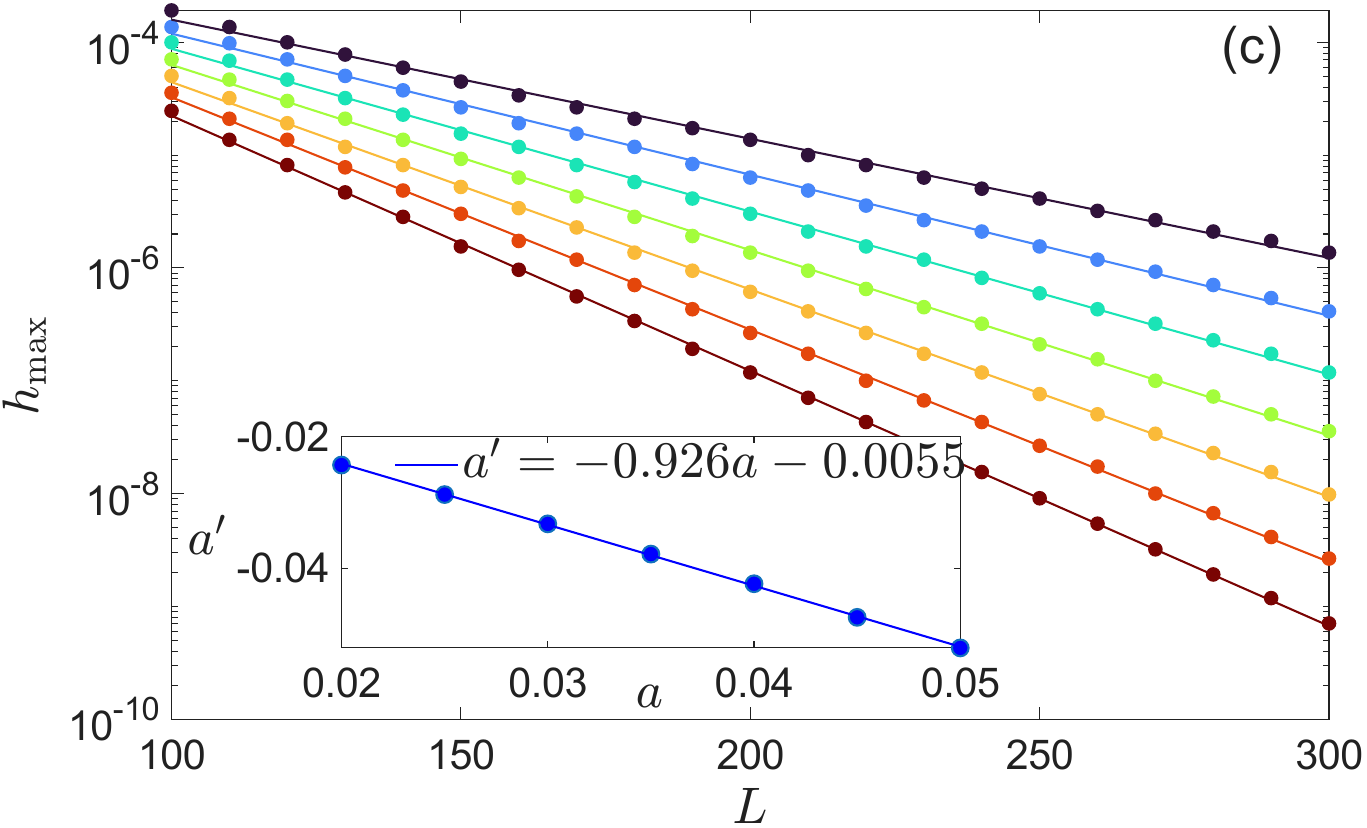}
    \caption{\textbf{Equilibrium Probe--SP--GS.} (a) QFI $\mathcal{F}_Q$ as a function of the field 
    strength $h$ for various probe sizes at fixed $a = 0.04$, when the single particle (SP) probe is prepared in the ground state (GS). 
    (b) Scaling of $\mathcal{F}_Q$ with system size $L$ evaluated 
    in the extended phase $h \to 0$, for several values of $a$. The numerical data (markers) are well described by the fitting function $\mathcal{F}_Q\propto e^{\beta L}$ (solid lines).
    Inset, the obtained exponents 
    $\beta({h \to 0}) = 1.62a + 0.0224$ and 
    $\beta({h \to h_{\max}}) = 1.829a + 0.0217$ extracted from fitting 
    exponent $\beta$ as a function of $a$ for both in the extended phase $({h \to 0})$ and at the transition point $({h \to h_{\max}})$.
    (c) Transition point $h_{\max}$ as a function of $L$ for several values of $a$. The transition point decays exponentially 
    with system size as $h_{\max} \propto e^{a'L}$, with fitted 
    exponents $a' = -0.926a - 0.0055$ shown in the inset as a function 
    of $a$. }
    \label{fig:SP_GS}
\end{figure}
To set the stage, we first revisit the single-particle Stark localization transition and the performance of the Stark probe for sensing a power-law field. 
Consider a one-dimensional probe consisting of $L$ sites hosting a single particle. The particle hops between nearest neighbors with amplitude $J=1$. In addition, the probe is subject to a gradient potential as $hV_j$. This system is described by a tight-binding Hamiltonian 
\begin{equation}\label{Eq:Stark_Hamiltonian}
H_{\rm{SP}} = -\sum_{j=1}^{L-1} \left( \vert j\rangle\langle j+1\vert + \vert j+1\rangle\langle j\vert  \right) + h\sum_{j=1}^{L}  V_{j} \vert j\rangle\langle j\vert.
\end{equation}
In this class of models, the single-particle spectrum is generically unstable to a gradient field, and irrespective of the specific functional form of $V_j$, an arbitrarily small field strength $h$ is sufficient to drive the system from an extended phase into a localized one throughout the full spectrum. 
Within the extended regime and also at the transition point, one finds that \emph{all} eigenstates display quantum-enhanced metrological scaling with the probe length $L$, as
$\mathcal{F}_{Q}\propto L^{\alpha \gamma+\beta} ,\quad \alpha>0,$
for power-law profiles $V_j=j^{\gamma}$~\cite{He2023PRL,Yousefjani2025PRApplied,Yousefjani2023CPB}. 
By contrast, sufficiently far inside the localized phase, this size-dependent enhancement disappears, the sensitivity ceases to grow with $L$ and instead saturates, approaching an almost universal behavior.  The boundary between these regimes is best understood within the framework of a continuous phase transition, marked by the appearance of a diverging length scale at criticality.
\\ \\
In this section, we investigate how replacing the conventional 
power-law gradient with an exponential potential $V_j = e^{aj}$ reshapes 
the sensing capabilities of Stark probes in the single-particle regime. In Fig.~\ref{fig:SP_GS} (a), we plot the 
ground-state QFI as a function of $h$ for probes of increasing size at 
fixed $a = 0.04$. The QFI displays a clear plateau over a finite range 
of $h$, which we identify as the extended phase, where the particle's 
wave function spreads across the chain with no preferred localization 
site. As $h$ increases, the system crosses into the localized phase 
through a transition at $h_{\max}$, marked by a sharp peak in the QFI. 
This transition point shifts toward smaller values as $L$ grows, 
approaching $h_{\max} \to 0$ in the thermodynamic limit $L \to \infty$, 
which means that an arbitrarily weak field eventually suffices to localize 
the probe. Beyond $h_{\max}$, the wave function collapses onto the sites 
with the lowest Zeeman energy and the QFI drops, first as 
$\mathcal{F}_Q \propto |h - h_{\max}|^{-2}$ and then more steeply 
as $\mathcal{F}_Q \propto |h - h_{\max}|^{-4}$ deeper inside the 
localized phase.
\\ \\
The central result of this section is the scaling behavior of the QFI across the extended phase and at the transition point. To make this precise,
we derive an analytical lower bound on the QFI by treating the field term
$hH_0$ as an infinitesimal perturbation to the integrable hopping
Hamiltonian $H_C$. This perturbative approach is justified precisely
because, for the single-particle system, localization sets in already at
$h \to 0$. Carrying out the calculation (see Appendix), the ground-state
QFI satisfies
\begin{equation}
    \mathcal{F}_Q > \frac{4e^{2a(L+1)}}{J^2(L+1)^2}\,\Theta(a),
    \label{eq:QFI_bound}
\end{equation}
where $\Theta(a)$ is a constant for fixed $a$ in the large-$L$ limit.
Obviously, sensitivity grows as $e^{2aL}$. This is a qualitative departure from anything achievable
with a power-law potential, where the best-known scaling is polynomial or
at most super-polynomial in $L$. We confirm this bound numerically in
Fig.~\ref{fig:SP_GS} (b), where the QFI extracted at $h \to 0$ for
probes of increasing size follow
\begin{equation}
    \mathcal{F}_Q \propto e^{\beta L},
    \label{eq:QFI_scaling}
\end{equation}
with fitted exponent
\begin{equation}
    \beta({h\to 0}) = 1.62a + 0.0224,
    \label{eq:beta_fits}
\end{equation}
reported in the inset. 
Repeating the same analysis for the obtained QFI in the transition point $h_{\max}$ results in 
\begin{equation}
    \beta({h\to h_{\max}}) = 1.829a + 0.0217.
\end{equation}
Results are presented in the inset of Fig.~\ref{fig:SP_GS} (b).
Both values are in good agreement with the $\beta \sim 2a$ dependence
predicted analytically. Notably, the obtained results confirm that the exponential enhancement
is not a fine-tuned feature, and it persists robustly
across the entire extended phase.
The transition point $h_{\max}$ marks the boundary between the extended 
and localized phases, and its behavior with system size carries important 
physical information. As shown in Fig.~\ref{fig:SP_GS}(c), for all values of 
$a$, the transition point decays exponentially with the chain length as
\begin{equation}
    h_{\max} \propto e^{a' L},  \quad \mathrm{with} \quad a' = -0.926a - 0.0055.
    \label{eq:hmax_scaling}
\end{equation}
This has a direct physical consequence, as the 
probe grows larger, an exponentially weaker field is sufficient to drive the system into the localized phase. In other words, larger probes are 
exponentially more sensitive to the gradient field, which is precisely 
what makes them powerful sensors. In the thermodynamic limit $L \to \infty$, 
the transition point vanishes, $h_{\max} \to 0$, confirming that the 
extended phase shrinks to a single point and the entire spectrum becomes 
localized for any nonzero $h$.
\\ \\
\begin{figure}[t!]
    \centering
    \includegraphics[width=\linewidth]{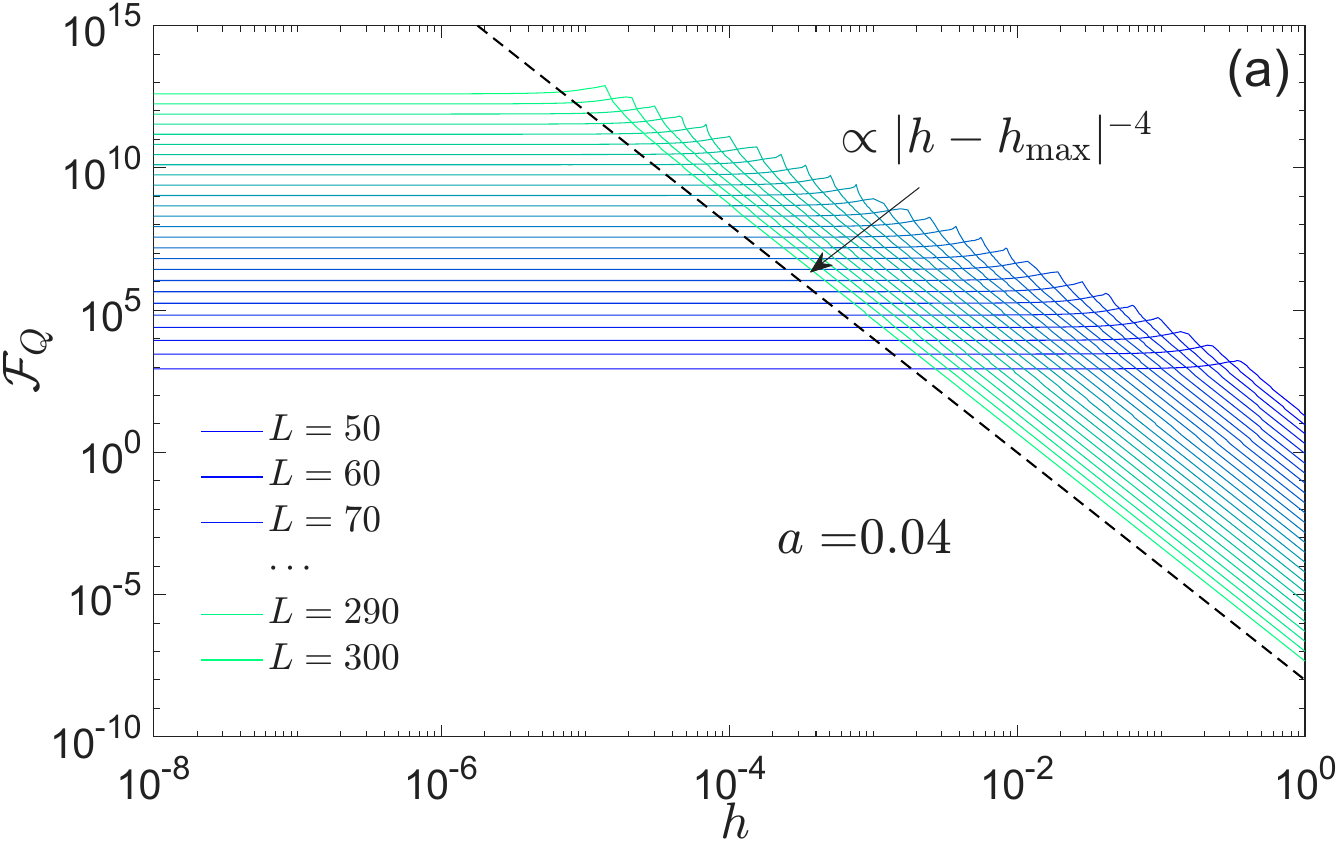}
    \includegraphics[width=\linewidth]{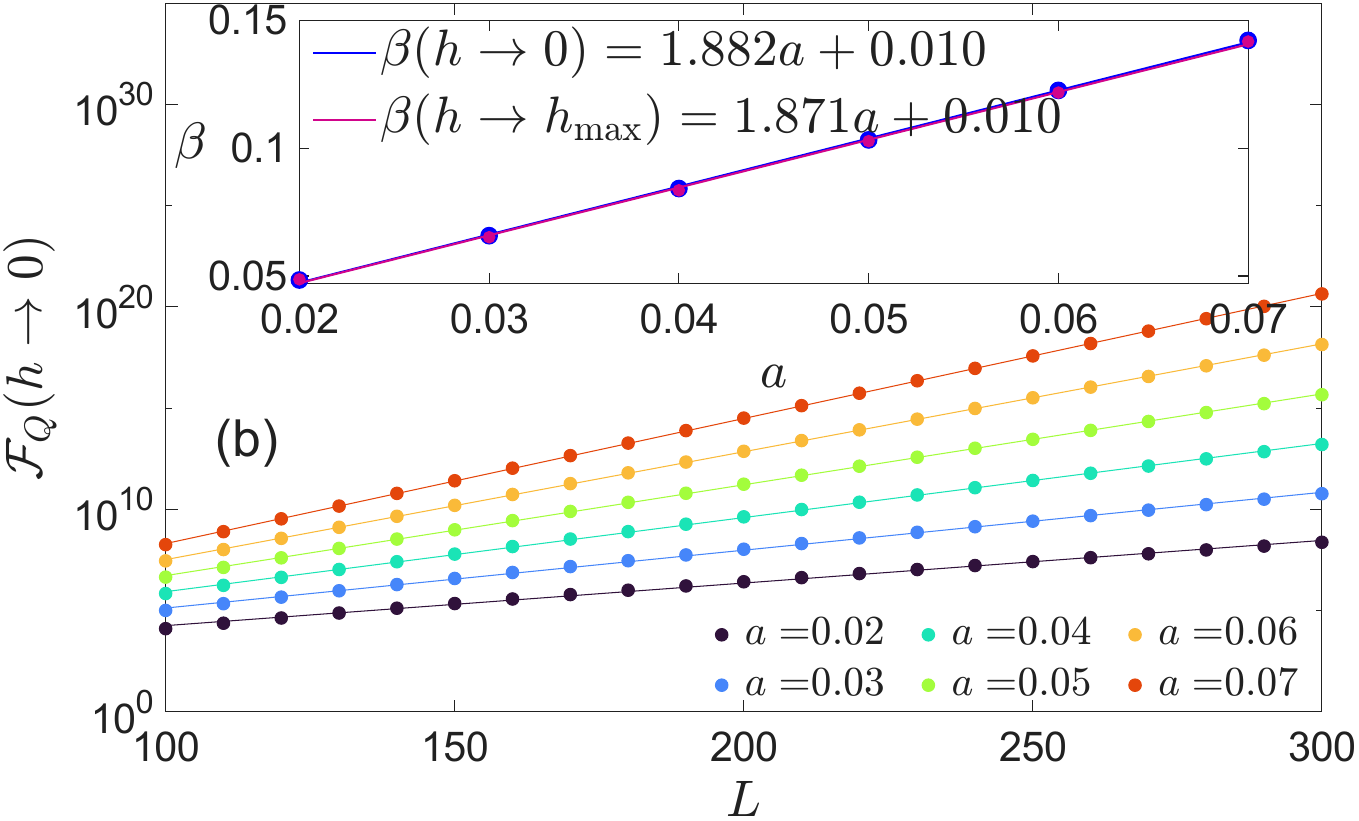}
    \includegraphics[width=\linewidth]{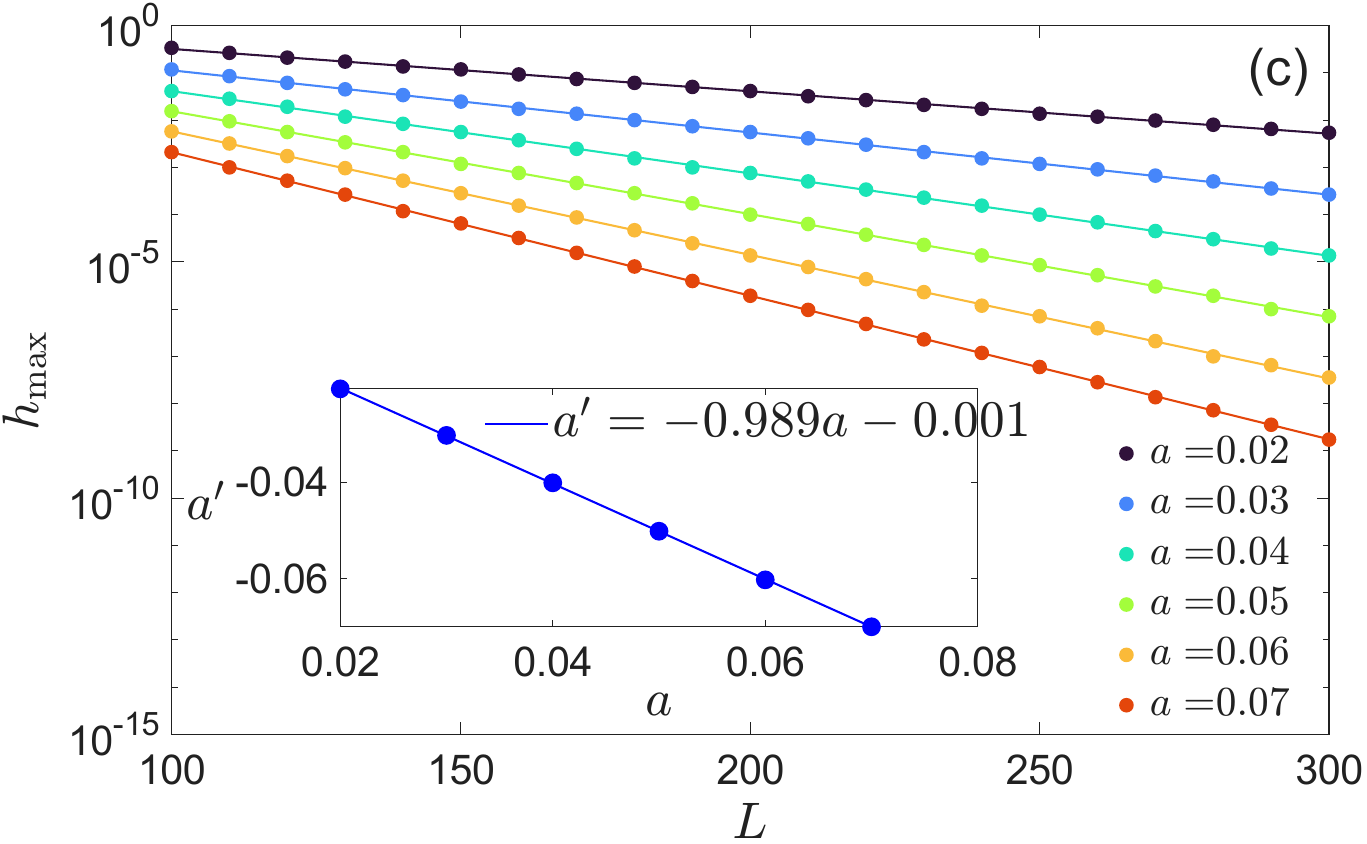}
    \caption{\textbf{Equilibrium probe--SP--MS.} (a) The QFI $\mathcal{F}_Q$ as a function of $h$ at 
    fixed $a = 0.04$ when single particle (SP) probe is prepared in a mid-spectrum (MS) eigenstate. Beyond the transition point $h_{\max}$, the QFI 
    decays as $\mathcal{F}_Q \propto |h - h_{\max}|^{-4}$.
    (b) Scaling of $\mathcal{F}_Q$ with system size $L$ evaluated 
    at $h \to 0$ and $h \to h_{\max}$ for several values of $a$, showing  $\mathcal{F}_Q\propto e^{\beta L}$ with 
    fitted exponents $\beta({h \to 0})= 1.882a + 0.010$ and 
    $\beta({h \to h_{\max}}) = 1.871a + 0.010$ reported in the inset.
    (c) Transition point $h_{\max}$ as a function of $L$ for 
    several values of $a$, with fitted exponents $a' = -0.989a - 0.001$ 
    shown in the inset.}
    \label{fig:SP_MS}
\end{figure}
Unlike most quantum sensors that achieve quantum enhancement only through the ground state, Stark-based probes offer a broader advantage, the exponential sensitivity enhancement is not tied to any particular energy 
level. To demonstrate this, we initialize the probe in the mid-spectrum 
eigenstate and compute the QFI as a function of $h$ for various systems 
sizes at fixed $a = 0.04$, with results shown in Fig.~\ref{fig:SP_MS} (a). 
The overall picture closely mirrors the ground state case. The QFI remains 
flat across the extended phase and develops a sharp peak at the transition 
point $h_{\max}$, beyond which it falls off as the wave function localizes. 
One notable difference from the ground state is the localization rate in 
the localized phase, where one finds $\mathcal{F}_Q \propto 
|h - h_{\max}|^{-4}$, a steeper decay compared to the $|h-h_{\max}|^{-2}$ 
behavior of the ground state. This faster decay reflects the higher 
sensitivity of mid-spectrum states to the onset of localization, and it 
signals a faster convergence to the thermodynamic limit deep inside the 
localized phase.
The scaling of the QFI in the extended phase and at the transition point 
is reported in Fig.~\ref{fig:SP_MS} (b). The numerical data are well described 
by the same exponential form $\mathcal{F}_Q \propto e^{\beta L}$, 
confirming that the exponential enhancement survives when the probe 
operates far from its ground state. The fitted exponents,
\begin{equation}
    \beta({h \to 0}) = 1.882a + 0.010, 
    \quad 
    \beta({h \to h_{\max}}) = 1.871a + 0.010,
    \label{eq:beta_MS}
\end{equation}
are slightly larger than their ground state counterparts, and strikingly, the two values are nearly identical. 
Finally, Fig.~\ref{fig:SP_MS} (c) shows the behavior of the transition point 
$h_{\max}$ as a function of $L$ for several values of $a$. As in 
the ground state case, the transition point decays exponentially with 
system size as $h_{\max} \propto e^{a'L}$, with fitted exponents 
$a' = -0.989a - 0.001$. This confirms that the exponential shrinking of the extended 
phase with system size is a robust feature of Stark probes, independent 
of which eigenstate is used to initialize the sensor.

\subsection{Many-body probe}
\begin{figure}[t!]
    \centering
    \includegraphics[width=\linewidth]{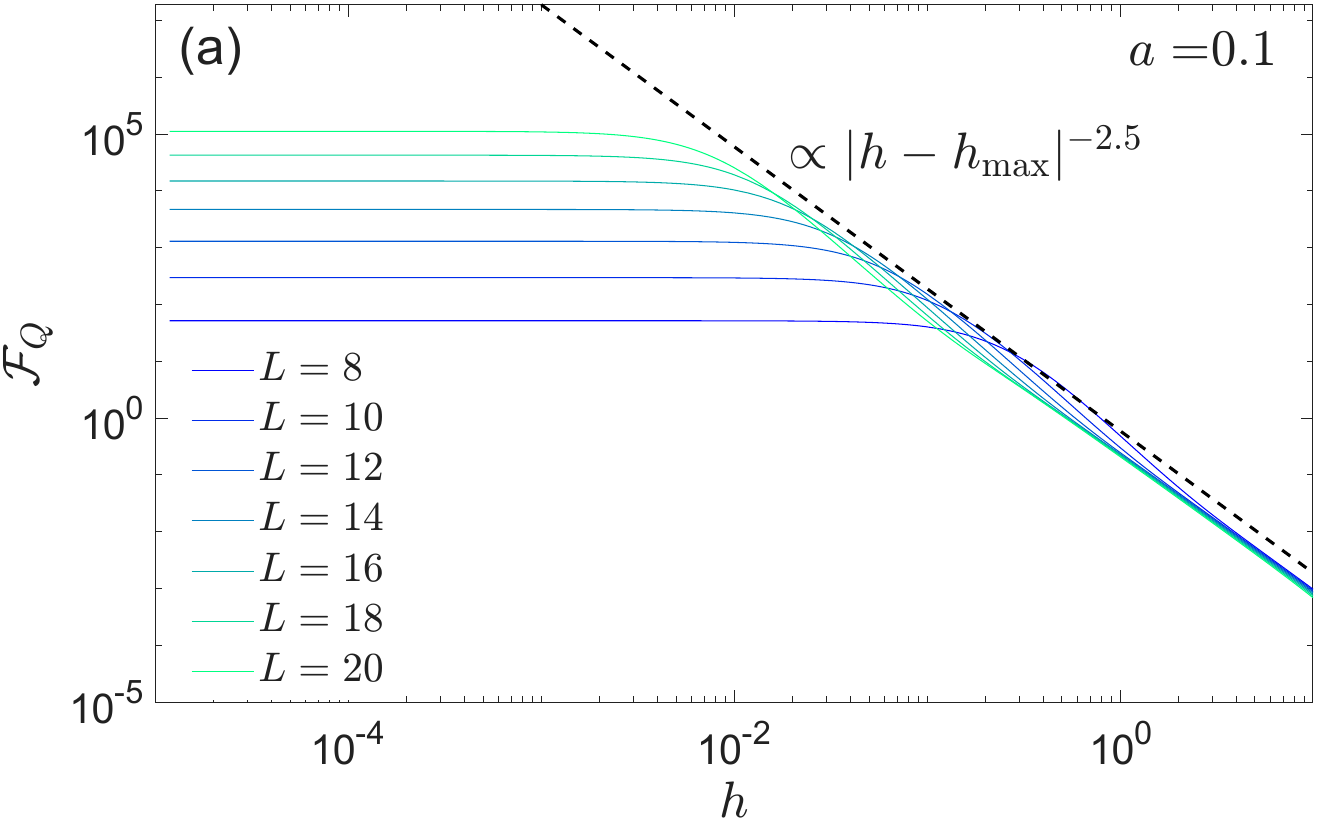}
    \includegraphics[width=\linewidth]{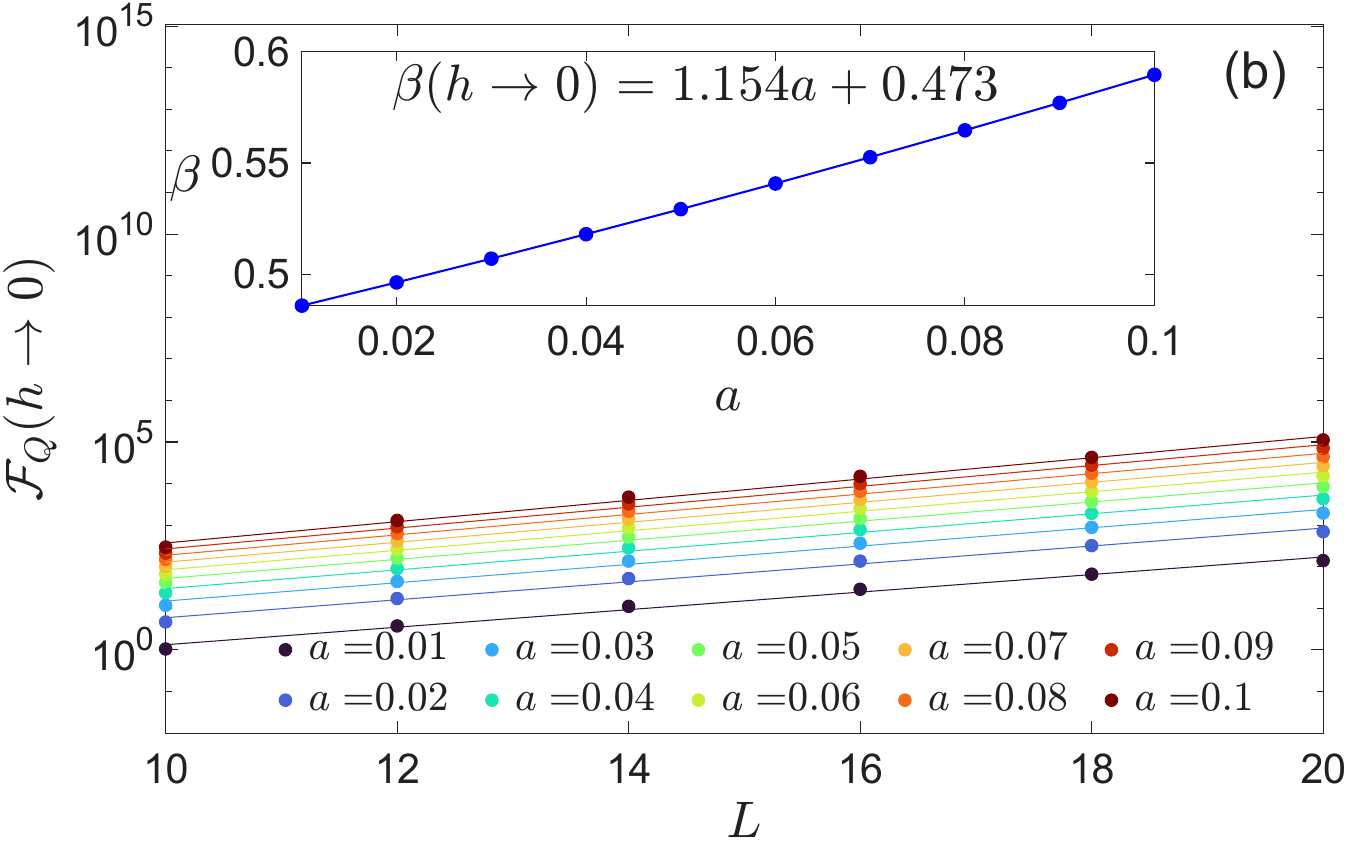}
    \caption{ \textbf{Equilibrium probe--MB.} (a) The QFI $\mathcal{F}_Q$ as a function of $h$ at 
    fixed $a = 0.1$, for many-body (MB) interacting probes of size $L = 8, 10, 12, \ldots, 20$ prepared in the ground state. 
    Beyond the transition point, the QFI decays as 
    $\mathcal{F}_Q \propto |h - h_{\max}|^{-2.5}$.
    (b) Scaling of $\mathcal{F}_Q$ with system size $L$ in the 
    ergodic phase ($h \to 0$) for several values of $a$. The data are 
    well described by $\mathcal{F}_Q \propto e^{\beta L}$, with 
    fitted exponents $\beta(h \to 0) = 1.154a + 0.473$ shown in the inset.}
    \label{fig:MB}
\end{figure}
Having established the exponential enhancement in the single-particle 
regime, we now ask whether many-body interactions preserve or destroy 
this advantage. 
We consider a one-dimensional chain of $L$ sites with $N = L/2$ 
interacting particles governed by the Hamiltonian
\begin{equation}
    H_{\rm{MB}} = -\sum_{j=1}^{L-1}\left(\sigma_j^x \sigma_{j+1}^x 
    + \sigma_j^y \sigma_{j+1}^y 
    + \sigma_j^z \sigma_{j+1}^z\right) 
    + h\sum_{j=1}^{L} e^{aj}\sigma_j^z,
    \label{eq:H_MB}
\end{equation}
where $\sigma_j^{x,y,z}$ are the Pauli operators, and $h$ is the strength of the exponential gradient potential 
$V_j = e^{aj}$. The ground-state QFI for sensing $h$ is presented in 
Fig.~\ref{fig:MB} (a) for fixed $a = 0.1$ and several system sizes. The 
qualitative structure is the same as in the single-particle case, a 
plateau in the ergodic phase followed by an algebraic reduction of QFI as the particles 
localize.
In Fig.~\ref{fig:MB}(b), we plot 
$\mathcal{F}_Q$ evaluated in the ergodic phase as a function of $L$ 
for several values of $a$. The numerical data are well described by
\begin{equation}
    \mathcal{F}_Q \propto e^{\beta L}, \quad \mathrm{with} \quad \beta(h\to 0)= 1.154a + 0.473
    \label{eq:QFI_MB_scaling}
\end{equation}
as shown in the inset 
of Fig.~\ref{fig:MB}(b). Comparing this directly with the single-particle case, two things stand out. First, the exponential enhancement survives 
the introduction of many-body interactions, confirming that Stark 
localization remains a powerful sensing resource beyond the single-particle 
level. Second, the constant offset in the exponent 
grows substantially, from values close to zero in the single-particle 
case to $0.473$ in the many-body case. This means that for any fixed $a$, the many-body probe achieves a parametrically larger QFI than its 
single-particle counterpart at the same system size. Rather than suppressing the sensing capability, interactions actively enhance it. 
This stands in sharp contrast to GHZ-based sensing protocols, where 
interactions during the sensing stage introduce uncontrolled phases that scramble the encoded information and degrade precision. 
\subsection{Resource analysis}
\begin{figure}[t!]
 \centering
    \includegraphics[width=\linewidth]{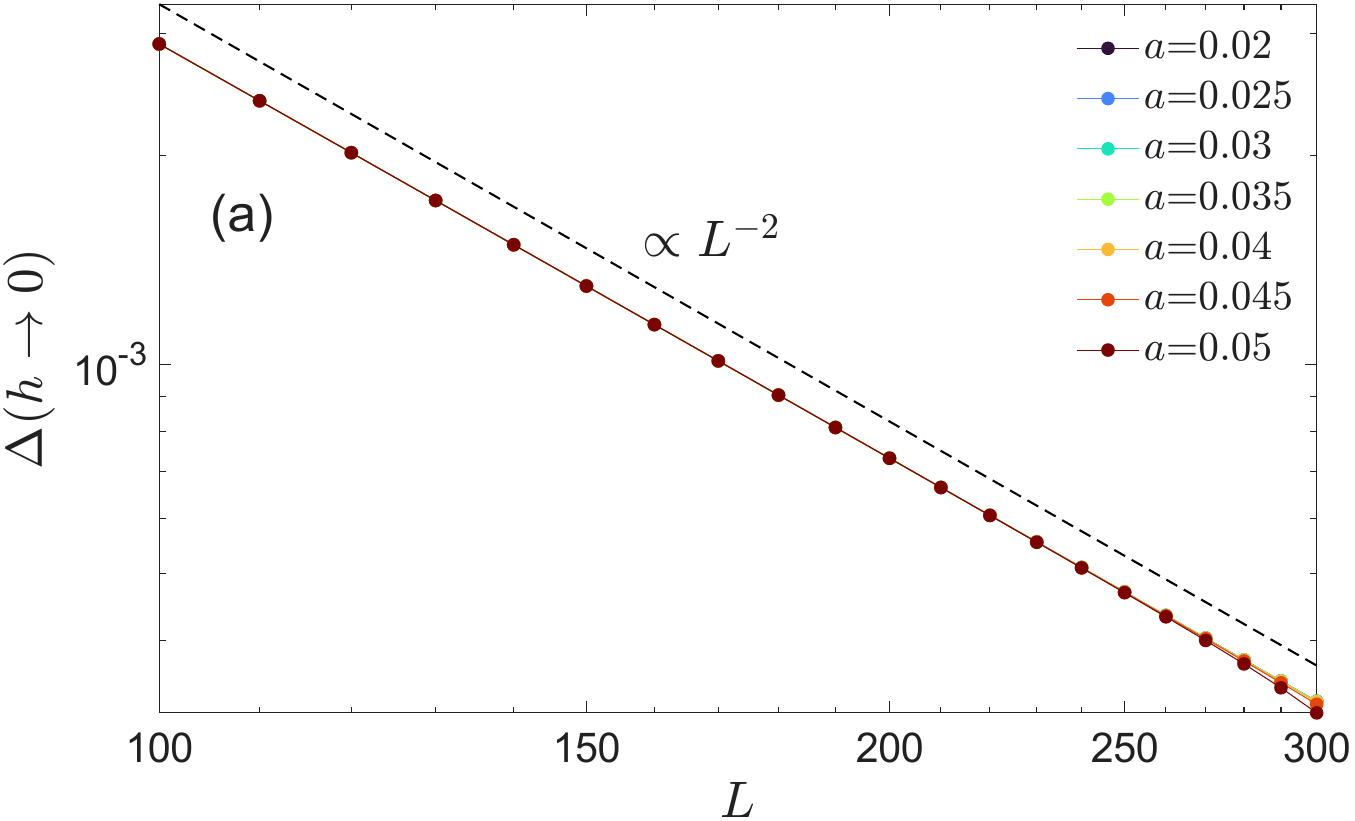}
    \includegraphics[width=\linewidth]{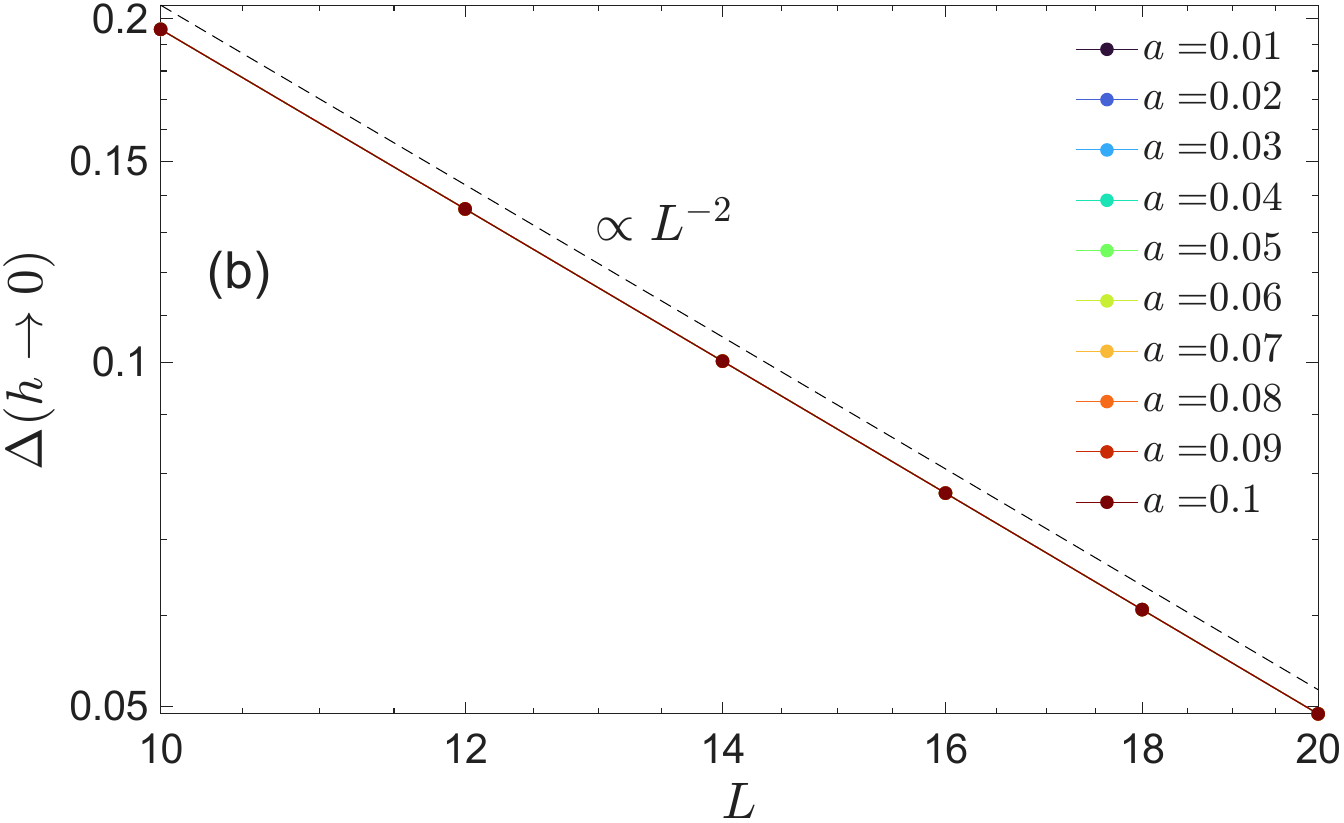}
    \caption{\textbf{Equilibrium probe.} Algebraic closing of the ground-state energy gap 
    $\Delta(h \to 0)$ as a function of system size $L$ for several 
    values of $a$, for the single-particle probe (a) and the many-body interacting probe (b).}
    \label{fig:Delta}   
\end{figure}
A complete assessment of any sensing scheme must account for all 
resources involved, not only the system size, but also the time required 
to prepare the probe state. For equilibrium-based probes, the ground state is typically prepared via adiabatic evolution, which requires a 
preparation time $\tau \sim 1/\Delta$, where $\Delta$ is the minimum energy 
gap of the Hamiltonian~\cite{McGeoch2014}. Incorporating this into the resource analysis, the relevant figure of merit is no longer the QFI 
alone, but the rescaled quantity $\mathcal{F}_Q/\tau$, which accounts for the fact that a longer preparation time reduces the number of measurements that can be performed within a fixed total time budget.
In Fig.~\ref{fig:Delta} (a) and (b), we plot the ground-state gap $\Delta$ as a function 
of $L$ in the extended phase ($h \to 0$) for both the single-particle 
and many-body probes, respectively. In both cases, the gap closes algebraically as
\begin{equation}
    \Delta(h \to 0) \propto L^{-2},
    \label{eq:gap_scaling}
\end{equation}
indicating that the preparation time grows only polynomially with 
system size as $\tau \sim L^2$. This is a crucial observation. Since the 
QFI scales exponentially as $\mathcal{F}_Q \propto e^{\beta L}$, the rescaled figure of merit becomes
\begin{equation}
    \frac{\mathcal{F}_Q}{\tau} \propto \frac{e^{\beta L}}{L^2},
    \label{eq:rescaled_QFI}
\end{equation}
which still grows exponentially with system size for large $L$. In other words, the polynomial cost of state preparation is entirely 
overwhelmed by the exponential gain in sensitivity, and the exponential quantum advantage survives the full resource analysis. 
This stands in sharp contrast to sensing schemes based on 
first-order phase transitions, where the energy gap closes 
exponentially as $\Delta \sim e^{-\alpha L}$, making the preparation time itself grow exponentially and potentially threatening the quantum advantage~\cite{sarkar2025exponentially}. In the Stark probe studied  here, the algebraic gap closing is therefore not merely a spectral 
property, but a key practical feature that keeps the sensing advantage 
intact when all resources are properly accounted for.

\section{Gradient field sensing; non-Equilibrium probe}
Having established the exponential sensing advantage of the equilibrium Stark probe in the previous section, we now turn to the non-equilibrium regime and ask whether the same advantage persists when the probe is no longer prepared in an eigenstate. Rather than relying on ground-state initialization, we consider a probe starting from a simple product state and evolving freely under the system's Hamiltonian. 
The unknown field $h$ is encoded entirely through the dynamical 
evolution, and the metrological performance is tracked via the 
time-dependent QFI in Eq.~(\ref{eq:QFI_dynamic}).

\subsection{Single-particle probe}
\begin{figure*}
    \centering
    \includegraphics[width=0.49\linewidth]{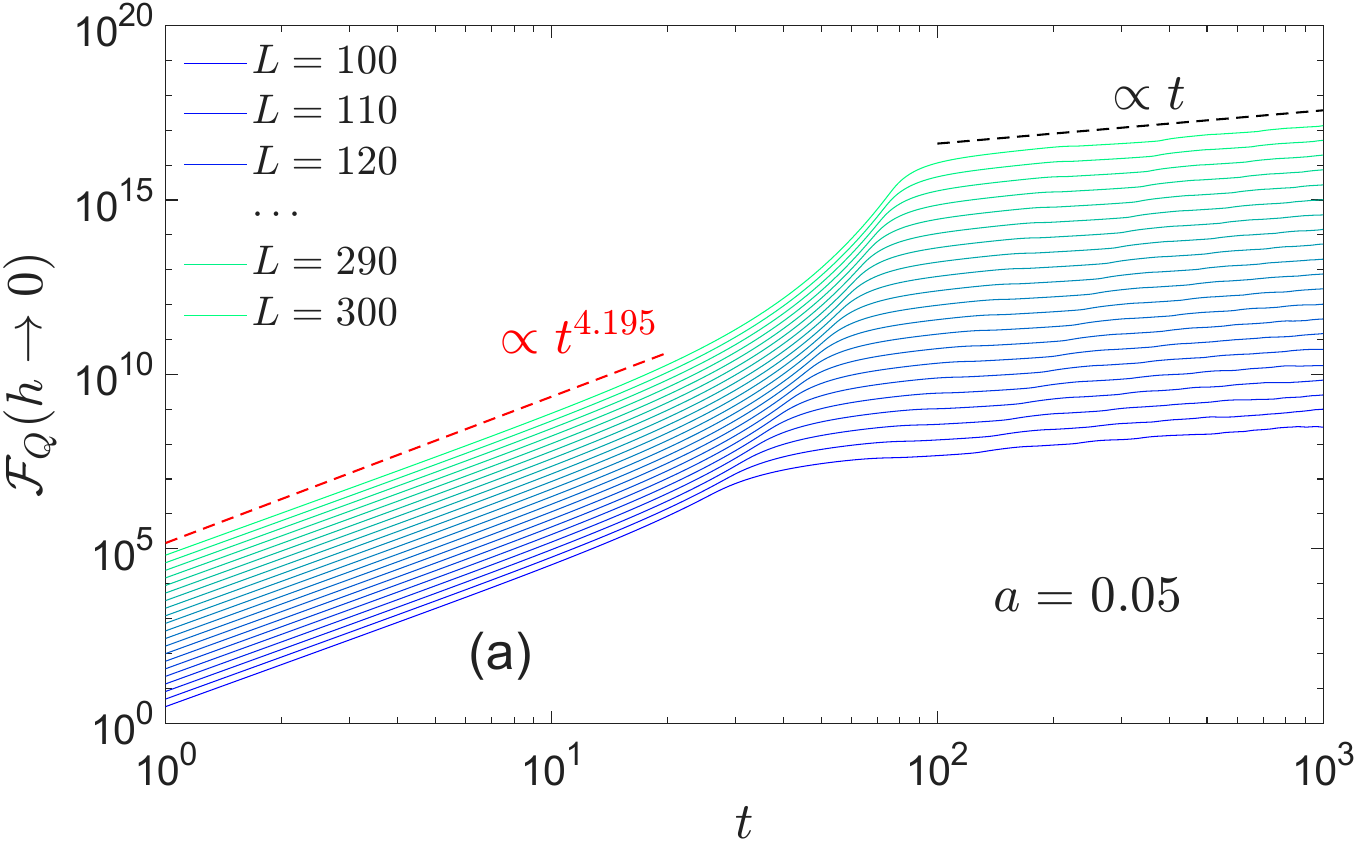}
    \includegraphics[width=0.49\linewidth]{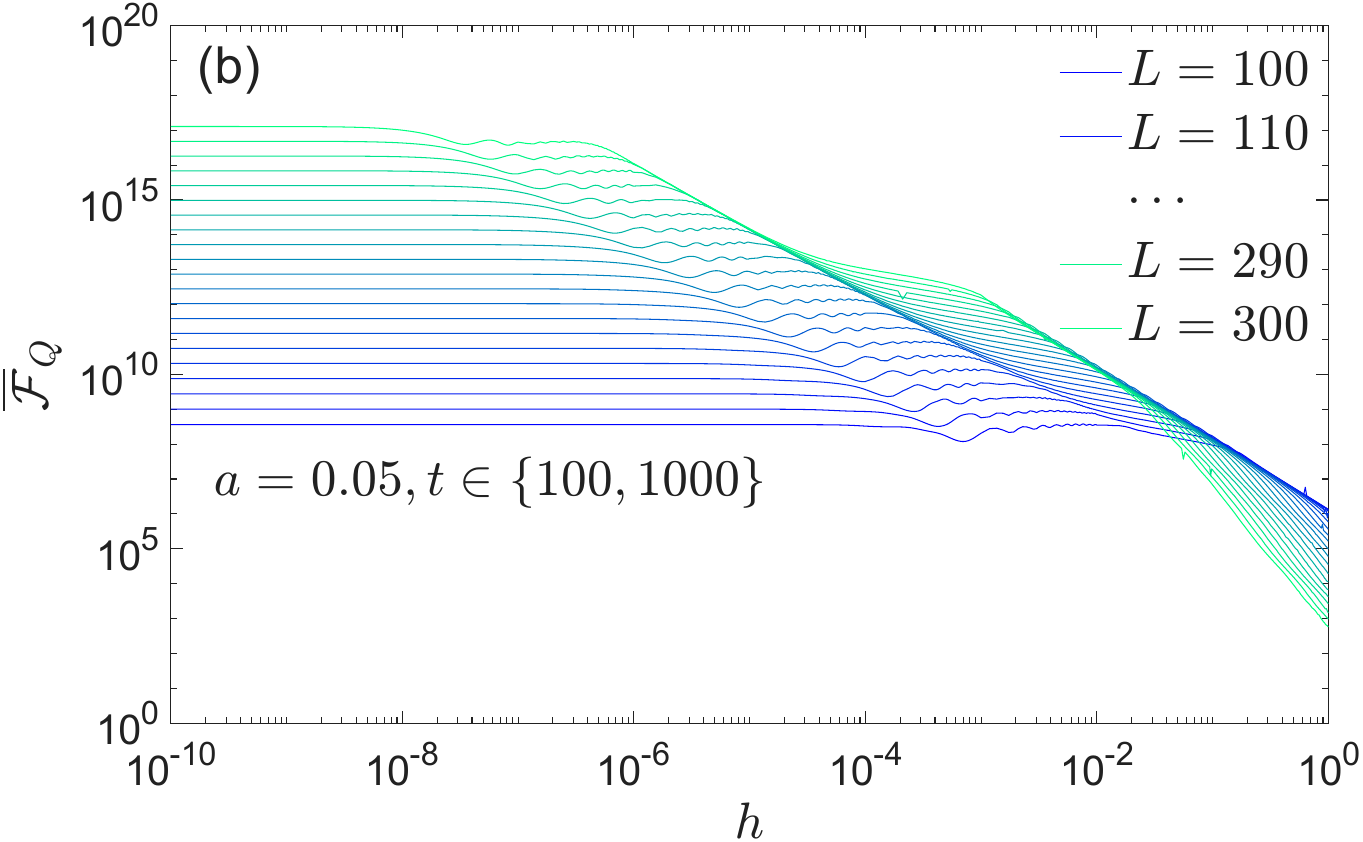}
    \includegraphics[width=0.49\linewidth]{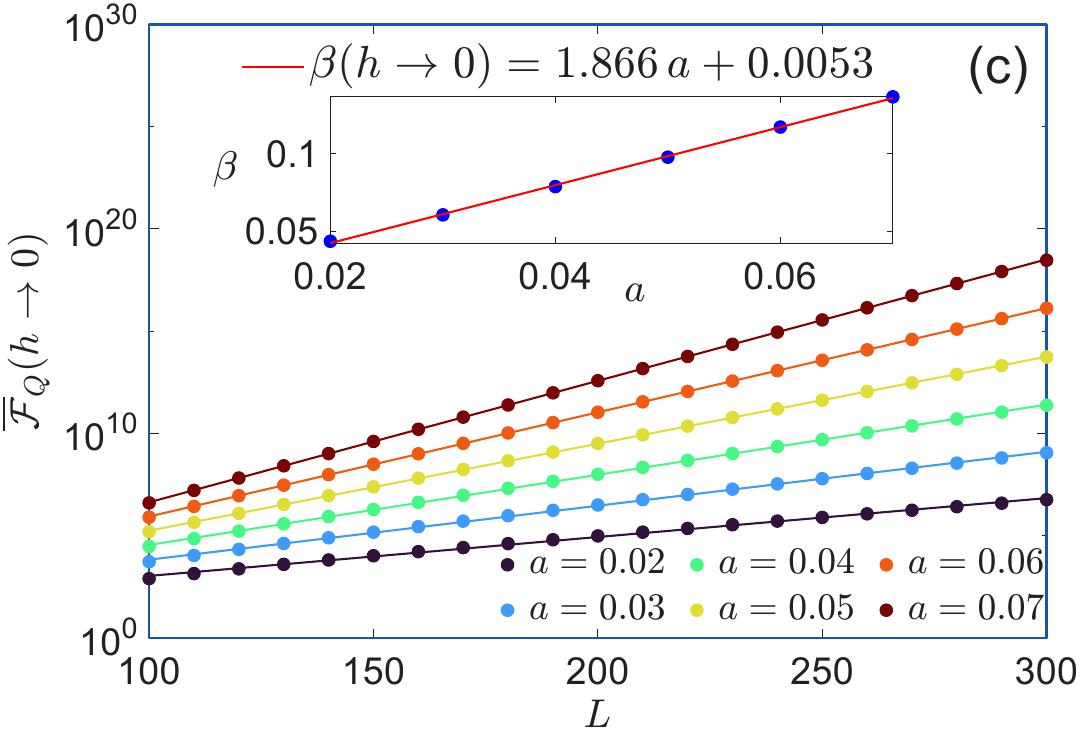}
    \includegraphics[width=0.49\linewidth]{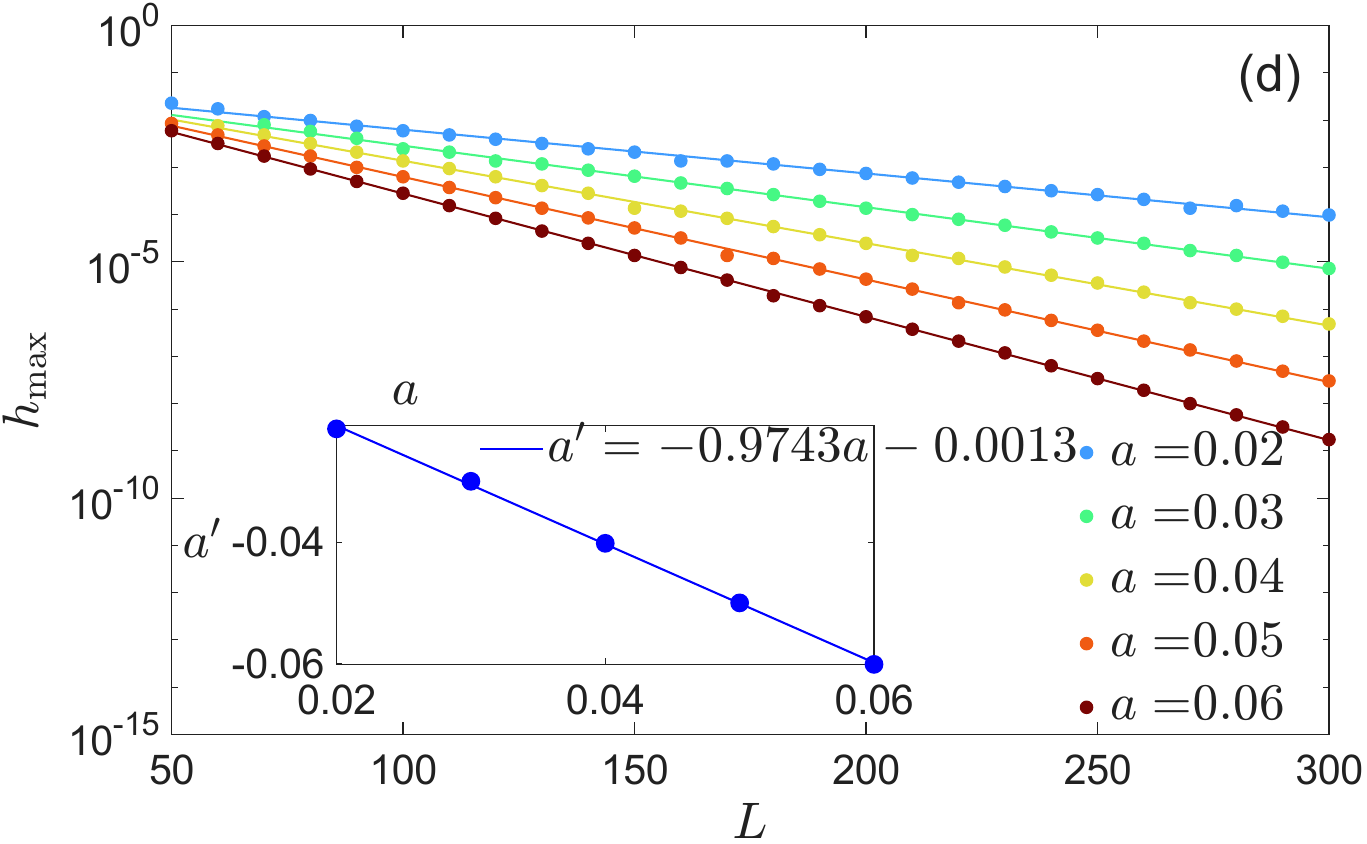}
        \caption{ \textbf{Non-equilibrium probe--SP.} (a) The dynamics of $\mathcal{F}_Q(h \to 0)$ for $a = 0.05$ and $L = 100, 110, \ldots, 300$ in single-particle (SP) probe initialized in a product state. The dashed lines indicate 
    $\mathcal{F}_Q \propto t^{4.195}$ and $\mathcal{F}_Q \propto t$. 
    (b) $\overline{\mathcal{F}}_Q$ as a function of $h$ for $a = 0.05$ 
    and $L = 100, 110, \ldots, 300$, averaged over $t \in [100, 1000]$. 
    (c) $\overline{\mathcal{F}}_Q(h \to 0)$ as a function of $L$ for 
    several values of $a$. Inset, the fitted exponents $\beta$ as a 
    function of $a$, with $\beta^{\rm Dy} = 1.866\,a + 0.0053$. 
    (d) Transition point $h_{\max}$ as a function of $L$ for several 
    values of $a$. Inset, the fitted exponents $a'$ as a function of 
    $a$, with $a' = -0.9743\,a - 0.0013$.}
    \label{fig:SP_Dy}
\end{figure*}
Unlike the equilibrium probe studied in the previous section, 
the non-equilibrium protocol requires no eigenstate preparation. 
Here, we initialize the probe in a simple product state, with the single excitation placed at the central site and 
subsequently allowed to evolve freely under the Hamiltonian in 
Eq.~(\ref{Eq:Stark_Hamiltonian}). The unknown field $h$ is then encoded in the time-evolved state $|\psi(t,h)\rangle = e^{-iH_{\rm{SP}}t}|\psi(0)\rangle$, 
and the metrological performance is quantified by the time-dependent QFI in Eq.~(\ref{eq:QFI_dynamic}).
We begin by examining the temporal behavior of the QFI deep inside the extended phase, shown in 
Fig.~\ref{fig:SP_Dy}(a) for $a = 0.05$ and probe sizes ranging from 
$L = 100$ to $L = 300$. Two distinct dynamical regimes are clearly 
visible. At short times, the QFI grows as a power law 
$\mathcal{F}_Q \propto t^{4.195}$, a super-quartic scaling that 
reflects the initial coherent spreading of the wave packet across 
the chain before it has had time to explore the full system. At 
longer times, this growth slows and crosses over into a linear 
regime $\mathcal{F}_Q \propto t$, signaling that the dynamics have 
entered a stationary scaling regime. In sharp contrast 
to the linear Stark case studied in 
Ref.~\cite{manshouri2025quantum}, the QFI curves for different system sizes are already separated at short times, before the long-time 
linear regime sets in. This reflects the exponential structure 
of the potential $V_j = e^{aj}$ as the energy scales 
exponentially across the chain, the dynamics feel the 
system size from the very beginning of the evolution.
\\ \\
To investigate the response of the system to the Stark field, one can focus on the saturation regime, namely $t \gtrsim 100$, visible in Fig.~\ref{fig:SP_Dy}(a).
To suppress residual oscillations within this regime, we define 
\begin{equation}
    \overline{\mathcal{F}}_Q 
    = \frac{1}{t_{\max} - t_{\min}} 
      \sum_{t = t_{\min}}^{t_{\max}} \mathcal{F}_Q(t),
    \label{eq:avQFI}
\end{equation}
with $t_{\min} = 100$ and $t_{\max} = 1000$, chosen to lie 
well inside the saturation window. This averaged quantity constitutes the figure of merit for the non-equilibrium probe, playing the 
same role as the ground-state QFI in the equilibrium analysis.
In Fig.~\ref{fig:SP_Dy}(b) we plot $\overline{\mathcal{F}}_Q$ as 
a function of $h$ for $a = 0.05$ and several system sizes. 
The overall structure closely mirrors the equilibrium phase 
diagram of Fig.~\ref{fig:SP_GS}(a). The averaged QFI is flat across the extended phase at small $h$, develops a peak at a transition point $h_{\max}$, and decays rapidly for 
$h > h_{\max}$ as the wave packet localizes. This confirms that the non-equilibrium dynamics faithfully inherit the phase 
structure of the underlying Hamiltonian, with the phase boundary encoded in the time-averaged signal, even though no eigenstate is ever prepared.
To extract the scaling of 
$\overline{\mathcal{F}}_Q$ with system size $L$, in 
Fig.~\ref{fig:SP_Dy}(c) we plot $\overline{\mathcal{F}}_Q$ 
evaluated in the extended phase ($h \to 0$) as a function of 
$L$ for several values of $a$. The data are well described 
by an exponential law,
\begin{equation}
    \overline{\mathcal{F}}_Q \propto e^{\beta L},
    \label{eq:dynamic_scaling}
\end{equation}
with fitted exponents shown in the inset and found to follow
\begin{equation}
    \beta(h\to 0) = 1.866\,a + 0.0053.
    \label{eq:beta_dynamic}
\end{equation}
Comparing this directly with the equilibrium single-particle 
result in Eq.~(\ref{eq:beta_fits}), the dynamic exponent  exceeds its 
equilibrium counterpart $\beta({h \to 0}) = 1.62\,a 
+ 0.0224$ for all studied values of $a$. This is a remarkable 
finding. Obviously, the non-equilibrium probe, initialized in a simple 
product state with no cooling or adiabatic ramp, achieves 
a parametrically larger exponential scaling exponent than the equilibrium probe operating at its ground state. 
This result can be understood qualitatively as follows. In the equilibrium setting, the QFI is determined entirely by 
the ground-state wave function. In the non-equilibrium setting, the initial product state has 
overlap with many eigenstates simultaneously, and the  subsequent unitary evolution coherently accumulates 
parameter-dependent phases from all of them. The information 
about $h$ is therefore encoded across the full spectrum 
rather than in a single eigenstate, which generically yields 
a larger sensitivity. The exponential potential $V_j = e^{aj}$ 
amplifies this advantage further because the exponentially 
large energy scales at the boundary sites dominate the 
parameter sensitivity, and these are sampled more 
efficiently by a delocalized initial state than by the 
ground state, which is biased toward the low-energy end of 
the chain.
Finally, Fig.~\ref{fig:SP_Dy}(d) shows the behavior of the 
transition point $h_{\max}$ extracted from the small peaks of 
$\overline{\mathcal{F}}_Q$ as a function of $L$, for 
several values of $a$. As in the equilibrium case, the 
transition point decays exponentially with the probe size,
\begin{equation}
    h_{\max} \propto e^{a' L}, 
    \quad a' = -0.9743\,a - 0.0013,
    \label{eq:hmax_dynamic}
\end{equation}
as shown in the inset of Fig.~\ref{fig:SP_Dy}(d). The fitted 
slope $|a'| \approx 0.974\,a$ is remarkably close to the 
equilibrium values $0.926\,a$ and $0.989\,a$ obtained for 
the ground and mid-spectrum states, 
respectively. This near-universal $|a'| \approx a$ scaling 
confirms that the exponential shrinking of the extended 
phase with system size is a structural property of the 
exponential Stark Hamiltonian itself, independent of 
whether the probe operates at equilibrium or far from it. 
In the thermodynamic limit $L \to \infty$, the transition 
point again vanishes, $h_{\max} \to 0$, so that the entire 
spectrum becomes localized for any nonzero field, and the 
exponentially enhanced sensitivity persists across the 
full extended phase without any fine-tuning of $h$.

\subsection{Many-body probe}
\begin{figure*}[t!]
    \centering
    \includegraphics[width=0.49\linewidth]{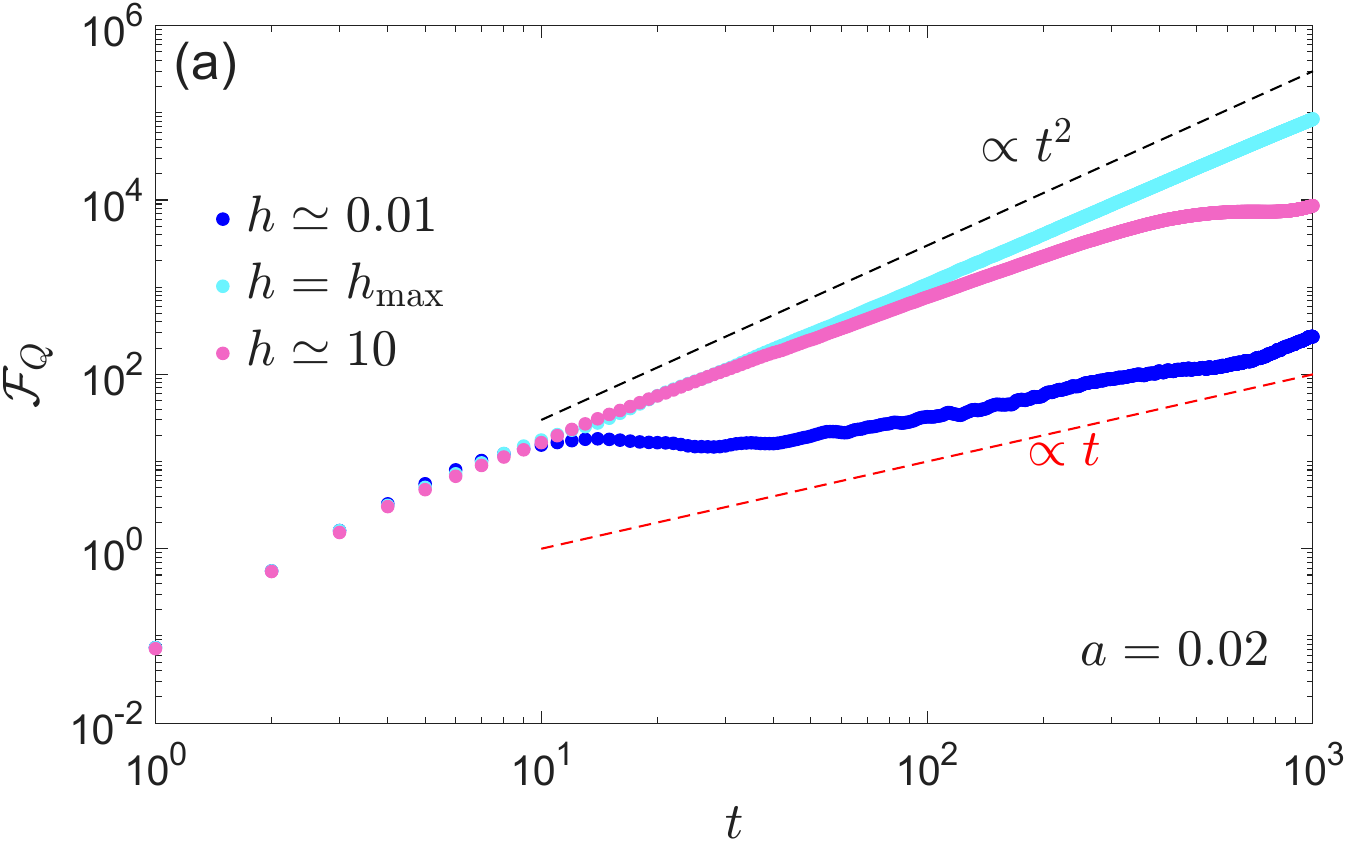}
    \includegraphics[width=0.49\linewidth]{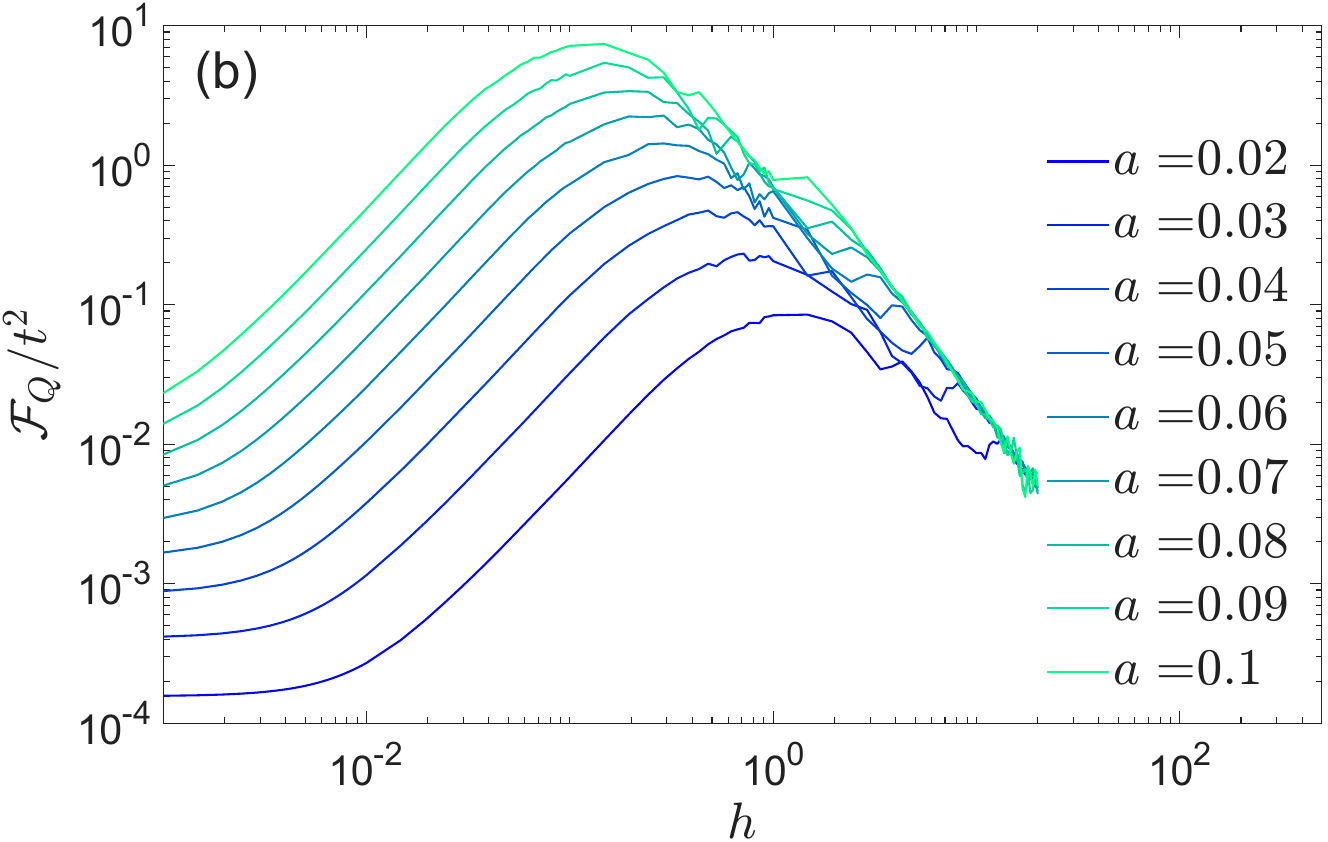}
    \includegraphics[width=0.49\linewidth]{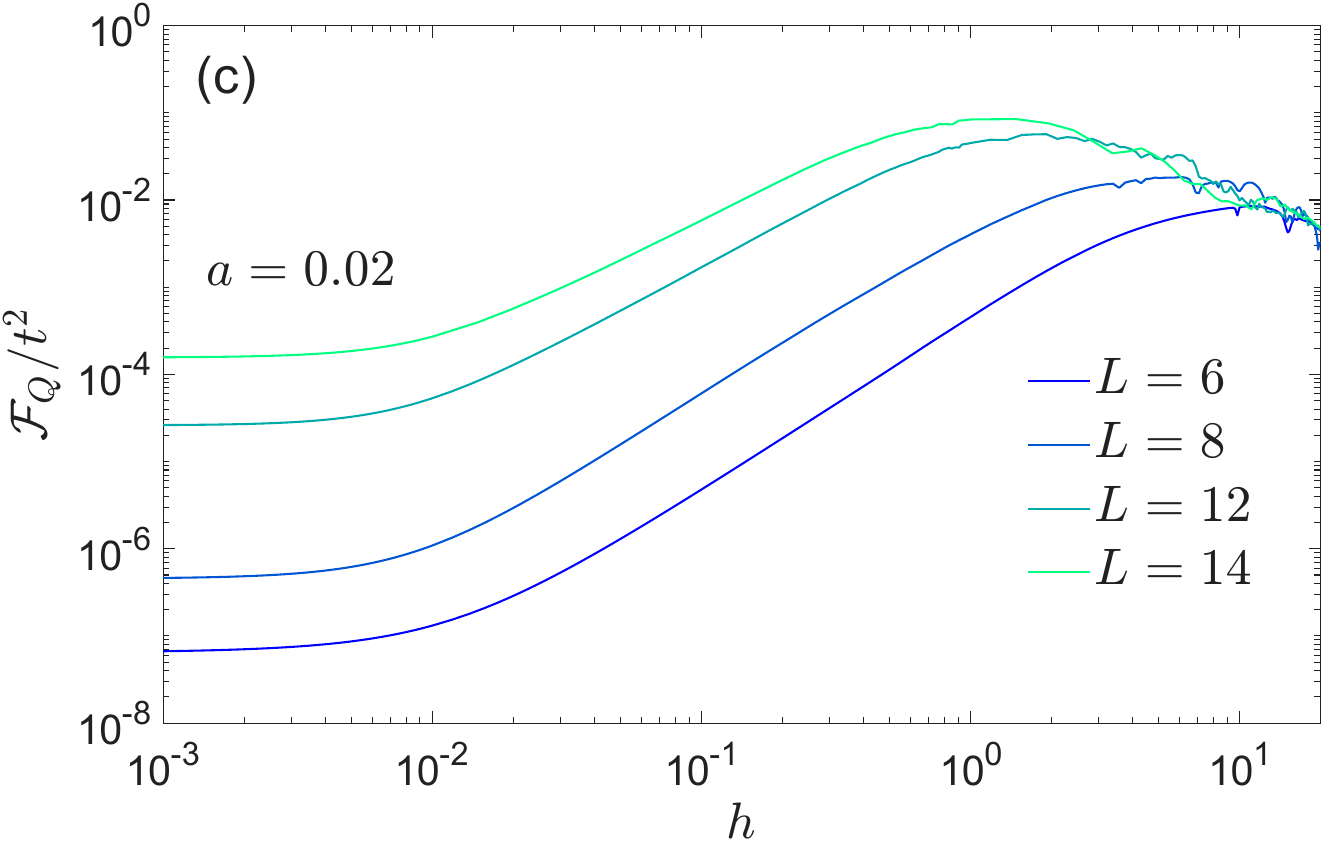}
    \includegraphics[width=0.49\linewidth]{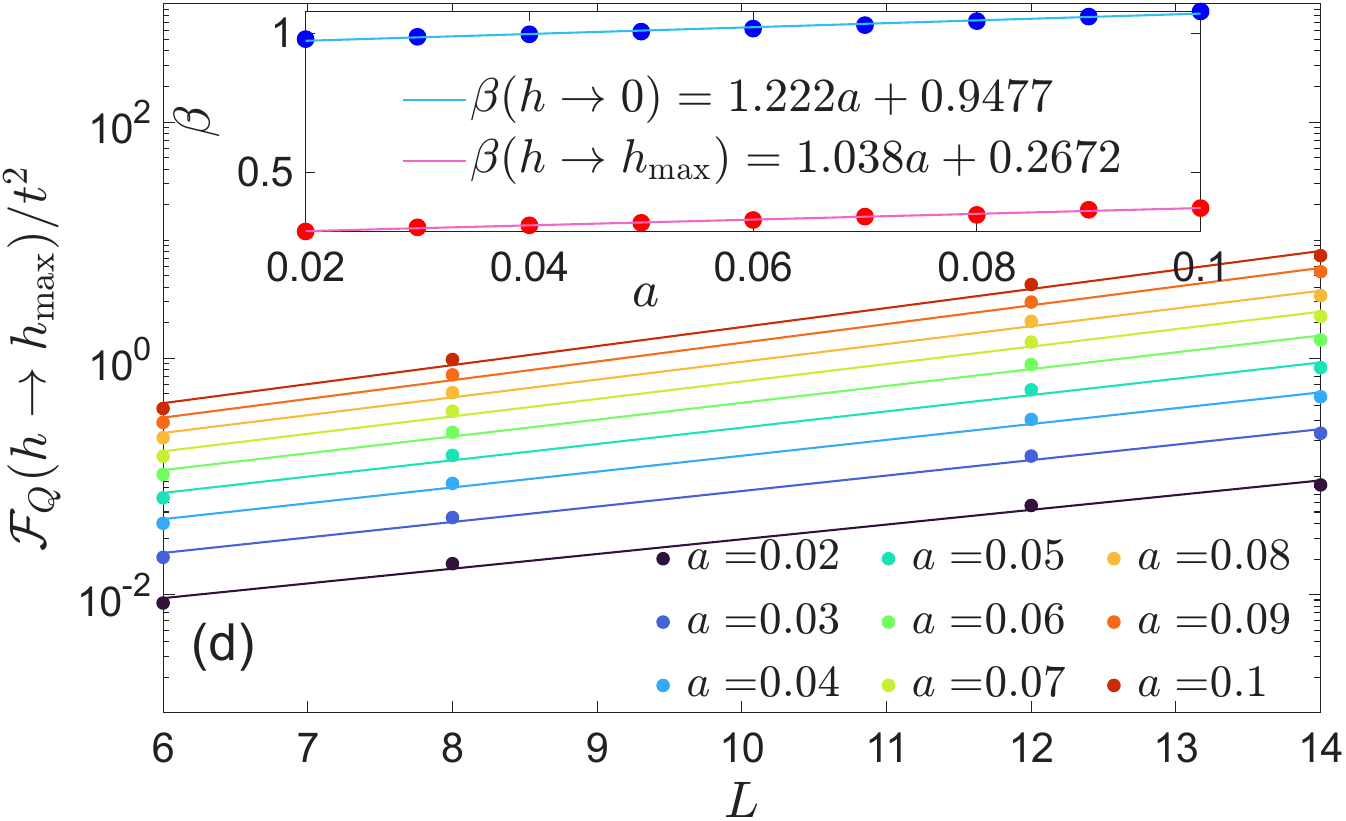}
    \caption{\textbf{Non-equilibrium probe--MB.} (a) The dynamics of $F_Q$ for
$a = 0.02$ when the many-body (MB) interacting probe of size $L{=}14$ is prepared in the N\'eel state. The three curves correspond to dynamics inside the ergodic phase ($h \simeq 0.01$),
at the transition point ($h = h_{\max}$), and in the localized phase ($h \simeq 10$). The dashed lines indicate
$F_Q \propto t^2$ and $F_Q \propto t$. (b) $F_Q/t^2$ as a function
of $h$ for several values of $a$ and fixed $L{=}14$. (c) $F_Q/t^2$ as a function of
$h$ at fixed $a = 0.02$ for $L = 6, 8, 12,$ and $14$. (d) Peak values of 
$F_Q(h \to h_{\max})/t^2$ as a function of $L$ for several values
of $a$. The numerical data (markers) are well fitted by the function $\mathcal{F}_Q\propto e^{\beta L}$. Inset, the fitted exponents $\beta$ as a function of $a$,
inside the ergodic phase with $\beta(h \to 0) = 1.222a + 0.9477$ and at the transition point with
$\beta(h \to h_{\max}) = 1.038a + 0.2672$.}
    \label{fig:Dy-MB}
\end{figure*}
Going beyond single particle probes, we now consider the many-body interacting chain introduced in Eq.~(\ref{eq:H_MB}), initialized in a simple half-filled N\'eel product state as~$|\psi(0)\rangle=|{\uparrow\downarrow\cdots\uparrow\downarrow}\rangle$,
and let it evolve freely under the many-body Hamiltonian $H_{\rm MB}$. The unknown field $h$ is encoded entirely through the unitary dynamics $|\psi(t,h)\rangle = e^{-iH_{\rm MB}t}|\psi(0)\rangle.$ 
Unlike the single-particle case, where a broad saturation window allowed us to define a time-averaged figure of merit, here the many-body response is more naturally characterized by the normalized quantity $F_Q/t^2$, since near the optimal working point the QFI itself approaches a quadratic growth in time.
In Fig.~\ref{fig:Dy-MB} (a), we show the time dependence of the QFI at fixed $a=0.02$ for three representative points, namely inside the ergodic phase ($h\simeq 0.01$), at the transition point ($h=h_{\max}$), and inside the localized phase ($h\simeq 10$). The dynamics reveal a clear hierarchy. Deep in the ergodic phase, the growth is relatively slow and is well captured by an approximately linear behavior $F_Q\propto t$ over the accessible time window. As the field is increased toward the transition point, however, the QFI is strongly amplified and approaches a quadratic growth $F_Q\propto t^2$, indicating that the many-body probe reaches Heisenberg scaling in time. For strong fields deep in the localized regime, the QFI remains appreciable but is clearly reduced compared with its value around $h_{\max}$. This identifies the transition region as the natural operating window of the dynamical many-body sensor.

To investigate the field dependence, in Fig.~\ref{fig:Dy-MB} (b), we plot $F_Q/t^2$ as a function of $h$ for several values of $a$ in a system of size $L=14$. For all cases, the normalized QFI develops a pronounced peak at $h_{\max}$, which separates the extended and localized regimes. The same qualitative behavior is observed in Fig.~\ref{fig:Dy-MB} (c), where $a=0.02$ is fixed, and the system size is varied. The peak height grows rapidly with $L$, showing that the many-body probe becomes more sensitive as the chain is enlarged. By contrast, sufficiently far inside the localized phase, the curves tend to collapse onto each other, indicating that the size dependence is gradually lost once the many-body wave function becomes localized.

The scaling analysis is summarized in Fig.~\ref{fig:Dy-MB} (d), showing the normalized QFI as a function of $L$ in the transition point $h\to h_{\max}$. Extracting the system-size dependence both in the ergodic phase ($h\to 0$) and at the transition point ($h\to h_{\max}$), we find that the normalized QFI is well described by
\begin{equation}
\frac{\mathcal{F}_Q}{t^2}\propto e^{\beta L},
\end{equation}
with fitted exponents
\begin{equation}
\beta({h\to 0})=1.222a+0.9477,
\quad
\beta({h\to h_{\max}})=1.038a+0.2672.
\end{equation}
These results show that the exponential size enhancement survives even in an interacting non-equilibrium many-body setting. In fact, the sizable positive offsets in the fitted exponents imply that the many-body probe remains highly sensitive even for modest values of $a$. Therefore, interactions do not destroy the metrological advantage of the exponential Stark profile. Rather, combined with a simple N\'eel state initialization and the absence of any cooling or adiabatic preparation stage, they provide a practically appealing route to exponentially enhanced weak-field sensing.

\section{Realization}
A superconducting realization of the exponential Stark sensing model requires a clear physical separation between the engineered spatial profile $e^{aj}$ and the unknown field amplitude $h$. We achieve this by engineering a site-dependent inductive susceptibility to a common unknown signal current, rather than programming the unknown parameter directly into the local qubit 
frequencies.

Consider a one-dimensional chain of $L$ flux-tunable transmon 
qubits with uniform nearest-neighbor hopping amplitude $J$, all 
inductively coupled to a common sensing bus carrying an unknown 
signal current $I_{\rm sig}$~\cite{Koch2007}. The physical origin 
of $I_{\rm sig}$ is a weak unknown magnetic flux threading a 
superconducting pickup coil connected in series with the sensing 
bus, which converts the unknown external flux into a proportional 
current. The flux threading the SQUID loop of qubit $j$ is then
\begin{equation}
    \Phi_j = M_j\, I_{\rm sig},
\end{equation}
where $M_j$ is the mutual inductance between the sensing bus and 
the SQUID loop of qubit $j$, which is a fixed, calibrated hardware 
parameter. The mutual inductances $M_j \propto e^{aj}$ can be realized by lithographically varying the local pickup geometry at each site, for example, through the coupling length, spacing, number of turns, or pickup-loop geometry, which together determine the bus-to-loop mutual inductance. 
The hopping amplitude $J$ is set by the capacitive or SQUID-mediated exchange interaction between neighboring transmons and depends only on the inter-qubit coupling geometry, which is 
fixed at fabrication. Adding a sensing flux through the SQUID 
loop shifts the on-site energy but leaves the inter-site coupling 
matrix elements unchanged, so $J$ and $h$ are structurally independent.
The required dynamic range $M_L/M_1 = 
e^{a(L-1)}$ is $e^{2} \approx 7.4$ for $a = 0.1$ and $L = 20$, which is well within the range achievable by standard on-chip inductive design. For larger chains, one has $e^5 \approx 150$ for $L = 50$ and $a = 0.1$, which 
demands careful layout but remains within the dynamic range of on-chip mutual inductances, which span from femtohenries to several nanohenries~\cite{Koch2007}. A gradiometric SQUID layout is preferable for the pickup elements, as it suppresses 
sensitivity to spatially uniform background fields while 
maintaining the desired nonuniform response to the signal 
current~\cite{Clarke2004}.
The main practical challenge is flux crosstalk, as the sensing bus and local flux-bias lines both couple inductively to multiple SQUID loops, the effective transfer matrix between currents and local fluxes must be characterized and compensated. 
This issue is well recognized in flux-tunable superconducting processors, and scalable crosstalk calibration protocols exist for transmon arrays~\cite{Krantz2019}. In the present proposal, the same methods can be applied to ensure that the effective coupling coefficients satisfy $M_j \propto e^{aj}$ after compensation for parasitic cross-couplings. Finally, state preparation and readout follow standard superconducting 
circuit protocols. 
State-of-the-art superconducting processors with $L \sim 50$--$100$ individually addressable qubits and single-qubit gate fidelities exceeding $99.9\%$ are already available~\cite{Krantz2019}, making this platform an immediate candidate for realizing the single-particle sensing protocol demonstrated in this work.

\section{Conclusion}
\begin{table*}[t]
\centering
\caption{The fitted exponential exponents $\beta$
characterizing the central size-scaling law $\mathcal{F}_Q\propto
e^{\beta L}$. }
\label{tab:beta_summary}
\renewcommand{\arraystretch}{1.35}
\begin{tabular}{|c|c|c|c|c|}
\hline
\textbf{Probe} & \textbf{System} & \textbf{Initial state} & $\boldsymbol{\beta(h\to 0)}$ & $\boldsymbol{\beta(h\to h_{\max})}$ \\
\hline

\multirow{3}{*}{Equilibrium}
& \multirow{2}{*}{Single-particle} & Ground state
& $1.62\,a + 0.0224$
& $1.829\,a + 0.0217$ \\
\cline{3-5}

&  & Mid-spectrum eigenstate
& $1.882\,a + 0.010$
& $1.871\,a + 0.010$ \\
\cline{2-5}

& Many-body & Ground state
& $1.154\,a + 0.473$
& -- \\
\hline

\multirow{2}{*}{Non-Equilibrium}
& Single-particle & Product state
& $1.866\,a + 0.0053$
& -- \\
\cline{2-5}

& Many-body & N\'eel state
& $1.222\,a + 0.9477$
& $1.038\,a + 0.2672$ \\
\hline
\end{tabular}
\end{table*}

We have shown that the spatial geography of the encoded field can serve as a metrological resource in its own right. By replacing the conventional linear or power-law Stark gradient with an exponential profile, $V_j=e^{aj}$, the sensing performance of Stark-localized probes is promoted from polynomial or super-polynomial scaling to a genuine exponential scaling with system size. In the equilibrium single-particle setting, this behavior is supported by an analytical lower bound on the quantum Fisher information and confirmed numerically across the extended phase and at the localization transition. The same qualitative picture survives beyond the ground state. The mid-spectrum eigenstates also display exponential enhancement, demonstrating that the advantage is not tied to a special energy level but is instead a structural consequence of the exponential Stark geometry. 
\\ \\ 
We further showed that this advantage is robust to interactions. For the interacting many-body probe, the QFI in the ergodic regime remains exponentially large in system size, with a larger fitted exponent than in the single-particle case, indicating that many-body effects do not suppress the sensing capability and can, in fact, enhance it. Just as importantly, this conclusion survives a fair resource analysis. Although the equilibrium protocol requires adiabatic state preparation, the relevant gap closes only algebraically, $\Delta(h\to 0)\propto L^{-2}$, so the preparation time grows only polynomially, $\tau\sim L^2$. As a result, the rescaled figure of merit $F_Q/\tau$ still grows exponentially, showing that the preparation overhead does not nullify the sensing advantage.
\\ \\
A particularly attractive outcome arises in the non-equilibrium protocol. Starting from a simple product state and encoding the unknown field through free unitary evolution, the relevant metrological figure of merit retains exponential scaling with system size, again of the form $\mathcal{F}_Q\propto e^{\beta L}$, with protocol-dependent exponent $\beta$ as summarized in Table~\ref{tab:beta_summary}. In particular, the single-particle dynamical probe exceeds the corresponding ground-state equilibrium exponent, while the many-body dynamical probe preserves exponential enhancement even in the presence of interactions. At the same time, the transition scale continues to shrink exponentially with system size, with $|a'|\simeq a$, just as in the equilibrium analysis. These results show that the exponential collapse of the extended phase is an intrinsic property of the exponential Stark Hamiltonian itself and not an artifact of a specific initialization scheme. The non-equilibrium route is especially compelling because it removes the need for cooling, adiabatic preparation,
and operation within a narrowly tuned sensing window while
preserving the exponential gain in precision.
\\ \\
Finally, we outlined a superconducting implementation in which the exponential Stark coupling is realized through graded mutual inductances between flux-tunable transmon
qubits and a common sensing bus. This proposal maintains a clean separation between the engineered spatial response and the unknown signal amplitude, making the sensing
architecture conceptually consistent as well as experimentally plausible. Taken together, our results identify exponentially graded Stark potentials as a distinct route to exponentially enhanced weak-field sensing, complementary to previously studied criticality-based mechanisms and promising both at equilibrium and far from it.

\section*{Acknowledgements}
The authors gratefully acknowledge Abolfazl Bayat for valuable scientific discussions and insightful comments.

\appendix
\setcounter{equation}{0}
\setcounter{figure}{0}
\setcounter{table}{0}
\renewcommand{\theequation}{A\arabic{equation}}
\renewcommand{\thefigure}{A\arabic{figure}}
\renewcommand{\thetable}{A\arabic{table}}
\begin{widetext}
\section*{Appendix: Theoretical analysis; single particle}
In the absence of the gradient field, namely $V_j = 0$, the model described in Eq.~(\ref{Eq:Stark_Hamiltonian}) becomes 
integrable and admits a straightforward diagonalization in terms of extended 
(Bloch-like) single-particle modes. The corresponding eigensystem reads
\begin{align}
    E_k = -2 \cos\!\left(\frac{k\pi}{L+1}\right), \qquad 
    |E_k\rangle = \sqrt{\frac{2}{L+1}} \sum_{j=1}^{L} (-1)^j 
    \sin\!\left(\frac{jk\pi}{L+1}\right) |j\rangle,
    \label{eq:eigensystem}
\end{align}
where $k = 1, \ldots, L$ labels the full set of eigenstates. These states 
have support over the entire chain and thus represent the extended phase. 
A key feature of the single-particle Hamiltonian is that the ground-state 
localization transition occurs already in the limit $h \to 0$. This enables 
a controlled description in terms of an integrable system subject to an 
infinitesimal perturbation. Building on this viewpoint, we develop an 
analytical approach to extract the scaling of the QFI. Starting from
\begin{equation}
    \mathcal{F}_Q = 4\sum_{k \neq 1} 
    \frac{|\langle E_k | H_0 | E_1 \rangle|^2}{|E_k - E_1|^2},
    \label{eq:QFI_sum}
\end{equation}
one obtains
\begin{equation}
    \mathcal{F}_Q = \frac{4}{(L+1)^2} \sum_{k \neq 1} \frac{N(k)}{D(k)}
    > \frac{4}{(L+1)^2} \frac{N(k=2)}{D(k=2)},
    \label{eq:QFI_bound_sum}
\end{equation}
in which
\begin{equation}
    N(k) = \left[\sum_{i=1}^{L} e^{aj}\sin(ik\theta)\sin(i\theta)\right]^2,
    \qquad
    D(k) = \left[\cos(k\theta) - \cos(\theta)\right]^2,
    \label{eq:ND}
\end{equation}
for $\theta = \pi/(L+1)$. The inequality in Eq.~\eqref{eq:QFI_bound_sum} 
follows because all terms $N(k)/D(k)$ are non-negative, and the $k=2$ term 
gives the dominant contribution among all $k \neq 1$.
In the following, we find a closed form for $N(k=2)/D(k=2)$. 
Using $\sin(2j\theta)\sin(j\theta) = \frac{1}{2}[\cos(j\theta) - \cos(3j\theta)]$, 
define
\begin{equation}
    C_L(\alpha) \equiv \sum_{j=1}^{L} e^{aj} \cos(j\alpha),
    \label{eq:CL_def}
\end{equation}
so that
\begin{equation}
    N(k=2) = \frac{1}{2}\left[C_L(\theta) - C_L(3\theta)\right]^2.
    \label{eq:Nk2}
\end{equation}
A standard geometric-series evaluation yields
\begin{equation}
    C_L(\alpha) = 
    \frac{e^a \cos\alpha - e^{2a} - e^{a(L+1)}\cos\!\left((L+1)\alpha\right) 
    + e^{a(L+2)}\cos(L\alpha)}
    {1 - 2e^a \cos\alpha + e^{2a}}.
    \label{eq:CL_exact}
\end{equation}
For $\alpha = \theta$ and $\alpha = 3\theta$, using $\theta = \pi/(L+1)$ one has
\begin{equation}
    \cos\!\left((L+1)\theta\right) = \cos\pi = -1, 
    \qquad 
    \cos(L\theta) = \cos(\pi - \theta) = -\cos\theta,
\end{equation}
and
\begin{equation}
    \cos\!\left((L+1)3\theta\right) = \cos 3\pi = -1, 
    \qquad 
    \cos(3L\theta) = \cos(3\pi - 3\theta) = -\cos(3\theta).
\end{equation}
These give the exact simplifications
\begin{align}
    C_L(\theta) &= \frac{e^a \cos\theta\,(1 - e^{a(L+1)}) 
    + (e^{a(L+1)} - e^{2a})}
    {1 - 2e^a\cos\theta + e^{2a}},
    \label{eq:CL_theta}\\[6pt]
    C_L(3\theta) &= \frac{e^a \cos(3\theta)\,(1 - e^{a(L+1)}) 
    + (e^{a(L+1)} - e^{2a})}
    {1 - 2e^a\cos(3\theta) + e^{2a}}.
    \label{eq:CL_3theta}
\end{align}
Since $e^a > 1$, the $e^{a(L+1)}$ terms dominate the numerators of 
both $C_L(\theta)$ and $C_L(3\theta)$ for large $L$. Extracting this 
leading contribution explicitly,
\begin{equation}
    C_L(\alpha) = e^{a(L+1)}\,
    \frac{1 - e^a \cos\alpha}{1 - 2e^a\cos\alpha + e^{2a}}
    + O(e^{2a}), 
    \qquad e^{a(L+1)} \gg e^{2a},
    \label{eq:CL_asymp}
\end{equation}
where the absolute remainder $O(e^{2a})$ is independent of $L$, the 
relative correction with respect to the leading term is 
$O(e^{-a(L-1)})$, which vanishes exponentially fast as $L \to \infty$.
Therefore, one has
\begin{equation}
    N(k=2) \sim \frac{e^{2a(L+1)}}{4}
    \left[
    \frac{1 - e^a\cos\theta}{1 - 2e^a\cos\theta + e^{2a}}
    -
    \frac{1 - e^a\cos(3\theta)}{1 - 2e^a\cos(3\theta) + e^{2a}}
    \right]^2,
    \label{eq:Nk2_asymp}
\end{equation}
and
\begin{equation}
    \frac{N(k=2)}{D(k=2)} \sim e^{2a(L+1)}\,\Theta(a,L),
    \label{eq:ratio_asymp}
\end{equation}
with
\begin{equation}
    \Theta(a,L) = 
    \left[
    \frac{1 - e^a\cos\theta}{1 - 2e^a\cos\theta + e^{2a}}
    -
    \frac{1 - e^a\cos(3\theta)}{1 - 2e^a\cos(3\theta) + e^{2a}}
    \right]^2
    \left[
    \frac{1}{\cos 2\theta - \cos\theta}
    \right]^2.
\end{equation}
This results in
\begin{equation}
    \mathcal{F}_Q > \frac{4e^{2a(L+1)}}{J^2(L+1)^2}\,\Theta(a,L).
    \label{eq:QFI_final}
\end{equation}
To show that $\Theta(a,L)$ converges to a finite constant as $L\to\infty$, 
we take $\theta \to 0$ and expand each factor to leading order in $\theta$. 
For the denominator factor, using $\cos(n\theta) \approx 1 - n^2\theta^2/2$,
\begin{equation}
    \cos 2\theta - \cos\theta 
    \approx \left(1 - 2\theta^2\right) - \left(1 - \frac{\theta^2}{2}\right) 
    = -\frac{3\theta^2}{2},
    \label{eq:denom_expand}
\end{equation}
so that
\begin{equation}
    \left[\frac{1}{\cos 2\theta - \cos\theta}\right]^2 
    \sim \frac{4}{9\theta^4}.
    \label{eq:denom_theta}
\end{equation}
For the numerator factor, applying $\cos(n\theta) \approx 1 - n^2\theta^2/2$ 
results in
\begin{equation}
    \left[
    \frac{1 - e^a\cos\theta}{1 - 2e^a\cos\theta + e^{2a}}
    -
    \frac{1 - e^a\cos(3\theta)}{1 - 2e^a\cos(3\theta) + e^{2a}}
    \right]^2
    \sim 
    \frac{64\,e^{2a}}{(1+e^a)^2(1-e^a)^6}\,\theta^4.
    \label{eq:numer_theta}
\end{equation}
The $\theta^4$ factors in Eqs.~\eqref{eq:denom_theta} 
and~\eqref{eq:numer_theta} cancel exactly in $\Theta(a,L)$, yielding a 
finite large-$L$ limit:
\begin{equation}
    \Theta(a,L) \;\xrightarrow{L\to\infty}\; 
    \frac{64}{9}\,\frac{e^{2a}(e^a+1)^2}{(e^a-1)^6}.
    \label{eq:Theta_limit}
\end{equation}
Therefore, for any fixed $a > 0$, $\Theta(a,L)$ tends to a finite constant 
as $L \to \infty$, which confirms that the exponential growth 
$\mathcal{F}_Q \propto e^{2aL}$ is the genuine asymptotic scaling of the 
ground-state QFI.
\end{widetext}

%

\end{document}